\documentclass[twocolumn]{jpsj2}

\def\runauthor{J. \textsc{Otsuki}}

\title{%
Excitonic Bound State in the Extended Anderson Model\\ with c-f Coulomb Interaction 
}

\author{%
Junya \textsc{Otsuki}\thanks{E-mail address: otsuki@cmpt.phys.tohoku.ac.jp}
}

\inst{%
Department of Physics, Tohoku University, Sendai 980-8578
}

\recdate{\today}

\abst{%
The Anderson model with the Coulomb interaction between the local and conduction electrons is studied in the semiconducting phase. 
Based on a perturbation theory from the atomic limit, leading contributions for the c-f Coulomb interaction are incorporated as a vertex correction to hybridization. 
An analytical solution shows that the effective attraction in the intermediate states leads to a bound state localized at the local electron site. 
Self-consistent equations are constructed as an extension of the non-crossing approximation (NCA) to include the vertex part yielding the bound state.
A numerical calculation demonstrates the excitonic bound state inside the semiconducting gap for single-particle excitations, and a discontinuity at the gap edge for magnetic excitations. 
}

\kword{%
exciton, bound state, mixed-valence semiconductor, c-f Coulomb repulsion, 
single-particle excitation, dynamical magnetic susceptibility, non-crossing approximation (NCA)
}

\begin{document}
\maketitle

\section{Introduction}

A class of materials called the Kondo semiconductor or the mixed-valence semiconductor has attracted much attention by its peculiar nature; 
its semiconducting behavior is characterized by a small energy gap emerging gradually with decreasing temperature. 
It indicates strong correlations among electrons as the origin of the energy gap. 

Typical compounds classified into this category include SmB$_6$, YbB$_{12}$ and TmSe.
One of noticeable features in the semiconducting phase is magnetic excitations depending strongly on temperature. 
Neutron scattering experiments on SmB$_6$ have shown that an inelastic peak appears around 14meV only in the semiconducting phase\cite{Alekseev}. 
With increasing temperature, the peak disappears corresponding to a transition to the metallic phase. 
Furthermore, the intensity of the excitation depends much on the wave-vector. 
The peak with strong temperature and wave-number dependences cannot be attributed to crystalline electric field excitations.
On the other hand, experiments on YbB$_{12}$ have shown several narrow inelastic peaks near the spin-gap edge\cite{Bouvet, Nefeodava, Mignot}. 
Intensity of one of the peaks increases strongly with decreasing temperature as in SmB$_6$. 

The above-mentioned magnetic excitation have been ascribed to an exciton, which here means a bound state localized around the rare-earth site\cite{Kikoin, Kasuya, Riseborough}. 
The exciton is originally a general concept in the optical absorption in semiconductors. 
Due to the long-range Coulomb interaction, the particle and hole excited by a photon attract with each other to form the hydrogenic bound state. 
As a result, the optical absorption takes place at discrete energies lower than the energy gap. 
Based on this concept, such a bound state due to the Coulomb interaction between the conduction electron and the rare-earth ion has been postulated in the mixed-valence systems. 
Namely, a conduction electron produced by mixing with a 4f electron are trapped around the rare-earth site by the effective attraction of the resultant 4f hole, and analogously a valence hole are trapped by the effective attractive potential around the 4f electron. 
Unlike the optical absorption, the excitonic bound state in this case can exist at the ground state, since the valence of the rare-earth ion may be continuously fluctuating through hybridization. 

Although the exciton has been invoked to interprete the characteristics of the semiconducting phase in SmB$_6$ and YbB$_{12}$, most descriptions in mixed-valence semiconductors are phenomenologically given on the assumption of the presence of the excitonic bound state. 
From the microscopic point of view, it is not obvious whether the bound state does appear in mixed-valence compounds.
The objective of this paper is to clarify an existence of the excitonic bound state in the mixed-valence semiconductors on the basis of a microscopic model. 


Since the excitonic bound state is due to a single-site effect, it is reasonable to consider a single rare-earth impurity embedded in a semiconductor. 
We restrict our attention to the single-impurity model. 
We consider the Coulomb repulsion between the 4f electron and conduction electrons as a source of the bound state, in addition to interactions in the conventional Anderson model.

Concerning the low-energy dynamics, it is known that an influence of the c-f Coulomb repulsion can be absorbed into other parameters. 
In fact, a numerical renormalization group (NRG) calculation has shown that the single-particle excitation spectrum computed with the c-f Coulomb term can be obtained from another set of parameters without it\cite{Costi-Hewson, Hewson}. 
No qualitative difference has arisen by the c-f Coulomb interaction in the metallic phase. 
However, the results cannot be applied simply to the semiconducting phase.

In this model, the Kondo effect occurs  provided the Kondo temperature exceeds the energy gap\cite{Saso, Takegahara}. 
On the other hand, the Kondo effect is suppressed if the energy gap is large compared to the Kondo temperature, on which we concentrate our attention. 
To construct an approximation which is proper in the semiconducting phase, we perform a perturbation expansion from the atomic limit. 
We shall take account of leading contributions for the c-f Coulomb interaction as a vertex correction to hybridization. 
Then, an appearance of the excitonic bound state will be demonstrated in the single-particle and the magnetic excitations. 

This paper is organized as follows. 
In the next section, we write the Anderson model with the c-f Coulomb repulsion in terms of $X$-operators, and give a brief summary of the resolvent method, by which we perform a perturbation analysis in the following section. 
Section 3 discusses a vertex part leading to the excitonic bound state. 
The binding energy is examined, in detail, for the on-site interaction in \S4. 
A self-consistent perturbation theory including the vertex part are developed as an extension of the NCA for the model with the c-f Coulomb interaction in \S5. 
Numerical results are given for the single-particle and the magnetic excitations in \S6.
We conclude with a summary in the final section.

\section{Model and the Resolvent Method}

We start from the Anderson model, which expresses a local nature of 4f electrons and strong correlations among them. 
The Anderson model is known to bear characteristics of the Kondo effect and the valence fluctuations\cite{Hewson}. 
We shall consider, in addition to the conventional Anderson model, the Coulomb repulsion between 4f electrons and conduction electrons to capture properties emerging from the local bound state. 

We assume the strong Coulomb repulsion between 4f electrons, and thus restrict 4f states to 4f$^n$ and 4f$^{n+1}$. 
In order to describe such a situation, we write the Hamiltonian in terms of $X$-operators defined by $X_{\gamma \gamma'}=|\gamma \rangle \langle \gamma'|$, where $|\gamma \rangle$ denotes 4f$^n$ and 4f$^{n+1}$ states.
On the other hand, $|\alpha \rangle$ stands for 4f$^n$ states, and $|\beta \rangle$ 4f$^{n+1}$ states hereafter. 
In this notation, the completeness of 4f states within the restricted Hilbert space of 4f$^n$ and 4f$^{n+1}$ is represented as
\begin{align}
	\sum_{\gamma} X_{\gamma \gamma} 
	= \sum_{\alpha} X_{\alpha \alpha} + \sum_{\beta} X_{\beta \beta} = 1.
\end{align}
With use of the $X$-operators, the Hamiltonian is written as
\begin{align}
	H &= H_0 + H_{\rm hyb} + H_{\rm cf},
\label{eq:Hamiltonian}
\end{align}
where
\begin{align}
	H_0 &= \sum_{\mib{k} m} \epsilon_{\mib{k}} c_{\mib{k}m}^{\dag} c_{\mib{k}m}
	 + E_{\gamma} X_{\gamma \gamma}, \\
	H_{\rm hyb} &= V\sum_{\mib{k}m \alpha \beta}
	 \langle \beta |f^{\dag}_m |\alpha \rangle X_{\beta\alpha} c_{\mib{k}m}
	+ {\rm h.c.}
\end{align}
Here $\epsilon_{\mib{k}}$ denotes the energy of conduction electrons with respect to the chemical potential, and bases of the conduction electrons are represented by the Wannier orbital centered at the impurity site. 
$E_{\gamma}$ is the energy level of 4f states $|\gamma \rangle$ in the atomic limit, and only their relative values are relevant. 
We take $V$ as a real constant independent of $\mib{k}$. 

The term $H_{\rm cf}$ in eq.~(\ref{eq:Hamiltonian}) represents the Coulomb repulsion between conduction electrons and 4f electrons, and is written as 
\begin{align}
	H_{\rm cf} = \sum_{\mib{kq}m} U_{\rm cf}(\mib{q}) c_{\mib{k}+\mib{q}m}^{\dag} c_{\mib{k}m}
	 (N_{\rm f} - \nu), 
\end{align}
where $N_{\rm f}$ is the 4f number operator. 
We have introduced a constant $\nu$, which produces a potential scattering independent of 4f electrons, to determine a standard for the c-f interaction. 
For 4f$^0$-4f$^1$ systems, $\nu$ is, in most cases, taken to be 0. 
In the general case of $n$, however, it is optional between $n$ and $n+1$. 
The term of $\nu$ gives rise to a change of the local density of states (DOS) of conduction electrons. 
Hence, we shall choose either $n$ or $n+1$ for the value of $\nu$ so as to largely cancel out the Hartree term. 
The wave-number dependence of the Coulomb interaction $U(\mib{q})$ is kept for an analytical analysis in \S\ref{sec:vf}. 
In constructing self-consistent equations in \S\ref{sec:scpt}, the interaction is eventually treated as a point contact as is often done in the actual evaluations. 
In the restricted Hilbert space of 4f$^n$-4f$^{n+1}$, $H_{\rm cf}$ is rewritten in terms of the $X$-operators as 
\begin{align}
	H_{\rm cf} &=\sum_{\mib{kq}m} U_1(\mib{q}) c_{\mib{k}+\mib{q}m}^{\dag} c_{\mib{k}m}
	 \sum_{\beta}X_{\beta \beta} \nonumber \\
	&\quad - \sum_{\mib{kq}m} U_0(\mib{q}) c_{\mib{k}+\mib{q}m}^{\dag} c_{\mib{k}m}
	 \sum_{\alpha} X_{\alpha \alpha},
\label{eq:cf_coulomb}
\end{align}
where we have introduced $\nu$-dependent coupling constant $U_1(\mib{q}) = (n+1-\nu)U_{\rm cf}(\mib{q})$ and $U_0(\mib{q}) = (\nu -n) U_{\rm cf}(\mib{q})$. 
\begin{figure}[t]
	\begin{center}
	\includegraphics[width=6cm]{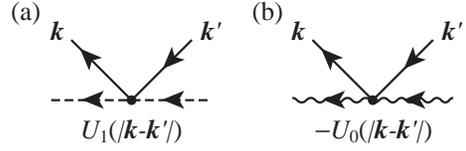}
	\end{center}
	\caption{Diagrammatical representations of the c-f Coulomb interaction corresponding to each term of eq.~(\ref{eq:cf_coulomb}). The solid, dashed and wavy line represent conduction electron, 4f$^{n+1}$ state and 4f$^n$ state, respectively}
	\label{fig:cf_coulomb}
\end{figure}
Figure~\ref{fig:cf_coulomb} shows diagrammatical representation of the c-f interaction.

In order to deal with the c-f Coulomb interaction, it is convenient to begin with the atomic limit. 
We can utilize conventional approaches developed for the Kondo model or the Coqblin-Schrieffer model.
Resolvent method is one of frameworks to perform the perturbation expansion from the atomic limit\cite{Keiter-Morandi, Bickers, Hewson, Kuramoto-K}.
A similar approach is also possible by using Abrikosov's pseudo-particle method\cite{Abrikosov, Read-Newns, Coleman}.
In this paper, we make use of the resolvent method.

The partition function $Z_{\rm f}$ of the 4f part is given in terms of the resolvent as
\begin{align}
	Z_{\rm f} = \int_{\rm C} \frac{{\rm d}z}{2\pi {\rm i}}
	 \text{e}^{-\beta z} \sum_{\gamma} R_{\gamma}(z),
\label{eq:part_func}
\end{align}
where the contour C encircles all singularities counterclockwise. 
The resolvent is given by
\begin{align}
	R_{\gamma}(z) = [z - E_{\gamma} -\Sigma_{\gamma} (z)]^{-1}.  
\label{eq:resolv}
\end{align}
The self-energy part $\Sigma_{\gamma}(z)$ is evaluated by the perturbation theory. 
The perturbation term is $H_{\rm hyb} + H_{\rm cf}$ in our model, eq.~(\ref{eq:Hamiltonian}).

Once the resolvents are obtained, physical quantities such as the single-particle excitations and the response functions are evaluated with use of them. 
Since expressions of physical quantities always involve the Boltzmann factor $\text{e}^{-\beta z}$, the resolvents are dealt with at real frequencies in actual computations.


We should note here that the term $H_{\rm hyb}$ can play a role similar to $H_{\rm cf}$. 
Namely, the second order process with respect to hybridization is regarded as an effective exchange or Coulomb-type interaction. 
The interaction naturally becomes antiferro-exchange or repulsive, as in the Kondo and the Coqblin-Schrieffer model derived from the Anderson model. 
We should therefore bear in mind that the repulsive c-f interaction may exist even though $H_{\rm cf}$ is missing. 
Such a situation will be demonstrated by numerical calculations in \S\ref{sec:results}.

\section{Intermediate State Interaction in Valence Fluctuations}
\label{sec:vf}
In the optical absorption in semiconductors or insulators, {\it final-state interaction}, that is, the Coulomb attraction between excited particle and hole, leads to the bound state called exciton, which is located in the energy gap. 
The relative motion of the particle-hole pair is described by the hydrogenic wave function with s-symmetry.
In the Green function method, such bound states are obtained by summing ladder-type diagrams in the perturbation series for the response function\cite{Mahan, Mahan_text}. 

Similar process is possible in the mixed-valence systems, where 4f states are continuously fluctuating between 4f$^n$ and 4f$^{n+1}$. 
Unlike the optical absorption, such a process takes place between the intermediate states of hybridization. 
Accordingly, the bound state should emerge into a variety of dynamical quantities related to the 4f electrons. 
To describe the dynamics, we begin with the NCA, which is the self-consistent perturbation theory with respect to hybridization\cite{nca1, Grewe}. 
The NCA gives a good description of the feature of the valence fluctuation apart from the c-f Coulomb interaction. 
The non-local c-f Coulomb interaction will be taken account of as a vertex correction to the NCA self-energy. 
We take, as the vertex part, the {\it intermediate-state interaction} of the ladder-type diagrams to derive an excitonic bound state, following refs.~\citen{Mahan} and \citen{Mahan_text}.

In this section, we omit the spin and orbital degeneracy for simplicity, since their degrees of freedom are not relevant to the following discussion on formation of the excitonic bound states.
We assume both 4f$^n$ and 4f$^{n+1}$ states to be singlet.

\subsection{Interaction between 4f$^{n+1}$ state and holes in valence band}
\begin{figure}[t]
	\begin{center}
	\includegraphics[width=8.5cm]{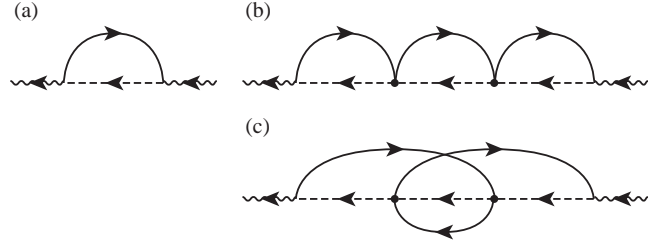}
	\end{center}
	\caption{Examples of the self-energy part $\Sigma_n(z)$ of 4f$^n$ state of order $V^2$. }
	\label{fig:self_example}
\end{figure}

We first consider the self-energy part $\Sigma_{n}(z)$ of the 4f$^n$ state. 
Figure~\ref{fig:self_example} shows typical diagrams of order $V^2$. 
Figure~\ref{fig:self_example}(a), which does not include the c-f Coulomb interaction, yields all diagrams without crossing of conduction lines when the resolvent is regarded as the renormalized one. 
In addition to the second order process with respect to hybridization, diagrams in Figs.~\ref{fig:self_example}(b) and (c) include interactions between 4f$^{n+1}$ intermediate state and the excited electron or hole. 
The solid lines going in the same direction as dashed line represent electron propagators, and correspond to the distribution function of the hole. 
On the other hand, lines with the opposite arrow stand for holes, and correspond to the electron distribution function. 
%
\begin{figure}[t]
	\begin{center}
	\includegraphics[width=6cm]{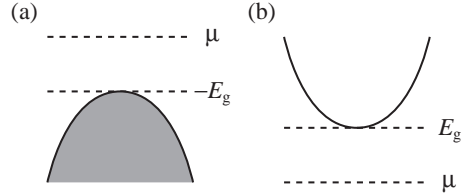}
	\end{center}
	\caption{Sketches of the (a) valence and (b) conduction band considered in subsection 3.1 and 3.2, respectively.}
	\label{fig:gap}
\end{figure}
We take almost filled valence band located $E_{\rm g}$ below the chemical potential as shown in Fig.~{\ref{fig:gap}(a). 
In this situation, a contribution from the diagram in Fig.~\ref{fig:self_example}(c) is negligible because of the hole distribution function, provided that temperature is much lower than $E_{\rm g}$. 
Consequently, the ladder-type diagrams like Fig.~\ref{fig:self_example}(b) give leading contribution among all intermediate state interactions in the semiconducting state.

The contribution from the ladder-type diagram modify the NCA self-energy as follows:
\begin{align}
	\Sigma_n(z) = V^2 \sum_{\mib{k}} f(\epsilon_{\mib{k}})
	 R_{n+1}(z+\epsilon_{\mib{k}}) \Lambda_{n+1} (\mib{k}, z), 
\label{eq:self_n}
\end{align}
where $f(\epsilon)$ is the Fermi distribution function. 
We have introduced the dimensionless vertex function $\Lambda_{n+1} (\mib{k}, z)$. 
The vertex part incorporates consecutive scattering of holes by 4f$^{n+1}$ state, and is given by the following integral equation: 
\begin{align}
	\Lambda_{n+1} (\mib{k}, z) = 1 &- \sum_{\mib{k}'} U_1 (|\mib{k}'-\mib{k}|) f(\epsilon_{\mib{k}'}) \nonumber \\
	&\quad \times R_{n+1}(z+\epsilon_{\mib{k}'}) \Lambda_{n+1} (\mib{k}', z).
\label{eq:vertex1}
\end{align}
Diagrammatical representations of eqs.~(\ref{eq:self_n}) and (\ref{eq:vertex1}) are shown in Fig.~\ref{fig:vertex1}(a) and (b), respectively. 
\begin{figure}[t]
	\begin{center}
	\includegraphics[width=8.5cm]{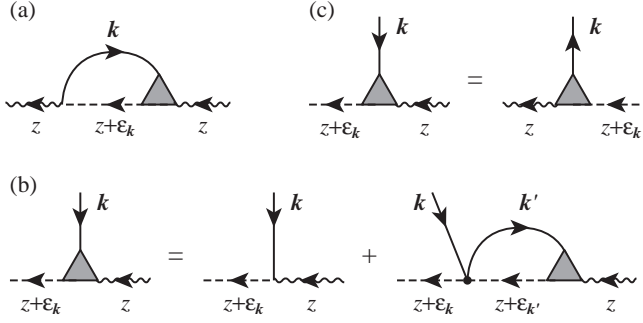}
	\end{center}
	\caption{Diagrammatical representation of (a) the self-energy part $\Sigma_n(z)$ of the 4f$^n$ state and (b) the equation of the vertex part $\Lambda_{n+1} (\mib{k}, z)$ corresponding to eq.~(\ref{eq:vertex1}). }
	\label{fig:vertex1}
\end{figure}
The minus sign of the second term in the right-hand side of eq.~(\ref{eq:vertex1}) comes from one permutation process of the conduction operators. 
It indicates the fact that the interaction between electron and hole is attractive. 
When $V$ depends on the wave vector $\mib{k}$, the following argument is also available by defining another vertex function $\tilde{\Lambda}_{n+1} (\mib{k}, z)$ with dimension of energy. 
Then an equation for $\tilde{\Lambda}_{n+1}(\mib{k}, z)$ is given by replacing the first term on the right-hand side of eq.~(\ref{eq:vertex1}) with $V_{\mib{k}}$. 

In Fig.~\ref{fig:vertex1}(a) and (b), we could represent the vertex part in a different manner as shown in Fig.~\ref{fig:vertex1}(c). 
Both diagrams express an identical function since now $V$ is independent of $\mib{k}$. 
We shall use both representations to construct a self-consistent scheme in \S\ref{sec:scpt}.

We solve eq.~(\ref{eq:vertex1}) assuming a simple situation, and then confirm that the vertex part introduced above can give a description of the excitonic bound state. 
We consider only the valence band with a parabolic energy dispersion of the form:
\begin{align}
	\epsilon_k = - E_{\rm g} -k^2/2m. 
\end{align}
The valence band is assumed to be completely filled, i.e., $f(\epsilon_k) = 1$,
which is valid at low temperatures such that $T\ll E_{\rm g}$ is satisfied.
Then, $R_{n+1}(z)$ can be replaced by bare one as follows:
\begin{align}
	R_{n+1}(z) \rightarrow R_{n+1}^{(0)}(z) = [z-\Delta]^{-1}, 
\end{align}
where $\Delta=E_{n+1} - E_n$ denotes an energy difference between 4f$^{n+1}$ and 4f$^n$ states. 
The above replacement is now exact since there is no unoccupied state to mix with 4f$^{n+1}$ state.
Correspondingly, eq.~(\ref{eq:self_n}) gives the exact self-energy. 

To derive the excitonic bound states, it is reasonable to deal with the equations in the real space. 
We introduce an auxiliary function $S(\mib{r}, z)$ defined by
\begin{align}
	S(\mib{r}, z) &= \sum_{\mib{k}} {\rm e}^{i\mib{k}\cdot \mib{r}}
	 R_{n+1}^{(0)}(z+\epsilon_k) \Lambda_{n+1}(\mib{k}, z) \nonumber \\
	&= \sum_{\mib{k}} {\rm e}^{i\mib{k}\cdot \mib{r}}
	 \frac{\Lambda_{n+1}(\mib{k}, z)}{z - \Delta - E_{\rm g} - k^2/2m},
\label{eq:def_S}
\end{align}
which is the Fourier transform of the integrand of eq.~(\ref{eq:self_n}).
The self-energy part $\Sigma_n(z)$ is represented with use of $S(\mib{r}, z)$ as follows:
\begin{align}
	\Sigma_n(z) = V^2 S(0, z).
	\label{eq:sigma_S}
\end{align}
Operating $(z - \Delta - E_{\rm g} + \nabla^2/2m)$ on eq.~(\ref{eq:def_S}), we obtain
\begin{align}
	[z - \Delta - E_{\rm g} + \nabla^2/2m] S(\mib{r}, z)
	 &= \sum_{\mib{k}} {\rm e}^{i\mib{k}\cdot \mib{r}} \Lambda_{n+1}(\mib{k}, z) \nonumber \\
	 &= \delta(\mib{r}) - U_1 (r) S(\mib{r}, z),
	 \label{eq:S}
\end{align}
where we have used eq.~(\ref{eq:vertex1}) in deriving the final expression. 
This equation includes the one-body Hamiltonian $H_{\rm ex}= -\nabla^2/2m - U_1 (r)$ with attractive potential, which can yield a bound state depending on $U_1 (r)$. 
The binding energy will be examined in detail for the short-range interaction in the next section. 
We represent eigenvalues of $H_{\rm ex}$ by $\epsilon_i$ as follows: 
\begin{align}
	H_{\rm ex} \phi_i(\mib{r}) = [- \nabla^2/2m - U_1 (r)]\phi_i(\mib{r}) = \epsilon_i \phi_i(\mib{r}).
\label{eq:Hamil_ex1}
\end{align}
We then expand $S(\mib{r}, z)$ in terms of the orthonormal bases $\phi_i(\mib{r})$. 
Solving eq.~(\ref{eq:S}) with respect to $S(\mib{r}, z)$, we obtain
\begin{align}
	S(\mib{r}, z) = \sum_i \frac{\phi_i^*(0) \phi_i(\mib{r})}{z - \Delta - (E_{\rm g} +\epsilon_i)}. 
\end{align}
This expression is substituted for eq.~(\ref{eq:sigma_S}) to yield
\begin{align}
	\Sigma_n(z) &= V^2 \sum_i \frac{|\phi_i(0)|^2}{z - \Delta - (E_{\rm g} +\epsilon_i)}
	 \nonumber \\
	&= V^2 \sum_i |\phi_i(0)|^2 R_{n+1}^{(0)}(z-E_{\rm g} - \epsilon_i).
	\label{eq:self_n_exciton}
\end{align}
This expression describes a virtual excitation of a hole, whose energy is modified by the intermediate-state interaction into $E_{\rm g} +\epsilon_i$. 
The energy $\epsilon_i$ given by eq.~(\ref{eq:Hamil_ex1}) may be negative due to the effective attraction, describing a bound state localized around the 4f site. 
Note that, because of the factor $|\phi_i(0)|^2$, only s-wave parts among $\phi_i(\mib{r})$ contribute to the self-energy. 
This results from the constant $V$, where a spread of the 4f electron is neglected.

\subsection{Interaction between 4f$^n$ state and electrons in conduction band}
The self-energy part $\Sigma_{n+1}(z)$ of 4f$^{n+1}$ state can be constructed in a manner similar to $\Sigma_n(z)$. 
In this case, interactions between 4f$^n$ intermediate state and excited conduction electrons are taken into account. 
In the ladder approximation, the NCA self-energy is altered into
\begin{align}
	\Sigma_{n+1}(z) = V^2 \sum_{\mib{k}} f(-\epsilon_{\mib{k}})
	 R_n (z-\epsilon_{\mib{k}}) \Lambda_{n} (\mib{k}, z), 
\label{eq:self_n+1}
\end{align}
where the vertex function $\Lambda_n (\mib{k}, z)$ is defined by
\begin{align}
	\Lambda_{n} (\mib{k}, z) = 1 &- \sum_{\mib{k}'} U_0 (|\mib{k}-\mib{k}'|) f(-\epsilon_{\mib{k}'})  \nonumber \\
	&\quad\times R_n(z-\epsilon_{\mib{k}'}) \Lambda_{n} (\mib{k}', z).
\label{eq:vertex0}
\end{align}
As in the $\Lambda_{n+1} (\mib{k}, z)$, the minus sign of the second term of the right-hand side indicates an attractive interaction. 
Corresponding diagrams are shown in Fig.~\ref{fig:vertex0}. 
The vertex part $\Lambda_{n} (\mib{k}, z)$ as well as $\Lambda_{n+1} (\mib{k}, z)$ can be expressed in a couple of ways as shown in the Fig.~\ref{fig:vertex0}(c).
\begin{figure}[t]
	\begin{center}
	\includegraphics[width=8.5cm]{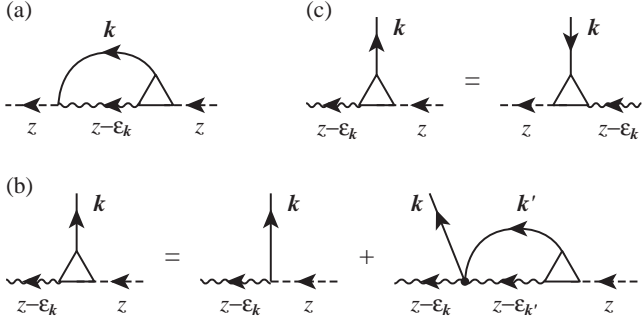}
	\end{center}
	\caption{Diagrammatical representation of (a) the self-energy part $\Sigma_{n+1}(z)$ of the 4f$^{n+1}$ state and (b) the equation of the vertex part $\Lambda_{n} (\mib{k}, z)$ corresponding to eq.~(\ref{eq:vertex0}). }
	\label{fig:vertex0}
\end{figure}

We solve eq.~(\ref{eq:vertex0}) for a toy model following the previous analysis. 
We suppose a conduction band with a parabolic dispersion represented by
\begin{align}
	\epsilon_k = E_{\rm g} +k^2/2m, 
\end{align}
which is sketched in Fig.~\ref{fig:gap}(b). 
We assume $T\ll E_{\rm g}$ so that the conduction band is taken to be completely empty, i.e., $f(-\epsilon_k) = 1$.
Then, $R_n(z)$ can be replaced by bare one
\begin{align}
	R_n(z) \rightarrow R_n^{(0)}(z) = z^{-1}, 
\end{align}
without any approximations. 
At the same time, the self-energy part $\Sigma_{n+1}(z)$ given by eq.~(\ref{eq:self_n+1}) becomes exact. 
On the above assumption, eq.~(\ref{eq:vertex0}) is solvable in the same procedure as the derivation of eq.~(\ref{eq:self_n_exciton}). 
We show only the result for $\Sigma_{n+1}(z)$:
\begin{align}
	\Sigma_{n+1}(z) &= V^2 \sum_i \frac{|\phi'_i(0)|^2}{z - (E_{\rm g} +\epsilon'_i)}
	\nonumber \\
	&= V^2 \sum_i |\phi'_i(0)|^2 R_n^{(0)}(z-E_{\rm g} - \epsilon'_i),
\label{eq:self_n+1_exciton}
\end{align}
where $\phi'_i(\mib{r})$ and $\epsilon'_i$ are determined by
\begin{align}
	H'_{\rm ex} \phi'_i(\mib{r}) = [- \nabla^2/2m - U_0 (r)]\phi'_i(\mib{r}) = \epsilon'_i \phi'_i(\mib{r}).
\label{eq:Hamil_ex0}
\end{align}
Equation~(\ref{eq:self_n+1_exciton}) indicates a virtual excitation of the electron with energy $E_{\rm g} + \epsilon'_i$, which may be smaller than $E_{\rm g}$ due to the effective attraction. 
Accordingly, it describes a situation that the 4f electron excited into the conduction band is bound around the impurity site.

\section{Binding Energy of the Local Bound State}
It has been shown that, in the absence of the thermal population of electrons or holes over the energy gap, virtual excitation processes are governed by the one-body Schr\"odinger equation with the attractive potential, eqs.~(\ref{eq:Hamil_ex1}) and (\ref{eq:Hamil_ex0}). 
It is apparent that the hydrogenic bound states appear provided the interaction extends to long range. 
In the metallic compounds, however, it is not clear because the interaction is short-ranged due to the strong screening effect. 
In order to see the condition for occurrence of the bound state, we consider an attractive $\delta$-function potential for each dimension
\begin{align}
	U(\mib{r})
	 = -U a^d \delta^d (\mib{r}), 
\end{align}
where $a$ denotes the lattice constant and $d$ is the spacial dimension.

From the Schr\"odinger equation with the attractive potential, we obtain the following condition which determines the binding energy $E_{\rm b}$:
\begin{align}
	1 = 2mU \frac1N \sum_{\mib{k}} \frac{1}{k^2 + \kappa^2}
	 = U\int {\rm d}\epsilon \rho(\epsilon) \frac{1}{\epsilon -E_{\rm b}},
\label{eq:bound_state_condition}
\end{align}
where $\kappa$ is defined by $\kappa=\sqrt{-2mE_{\rm b}}$ and $N$ denotes a number of sites.
Since we are now interested only in the bound state, we have restricted $E_{\rm b}$ to negative. 
We introduce a cutoff energy $D$ and corresponding wave number $k_{\rm c}=\sqrt{2mD}$ so as to satisfy the relation
\begin{align}
	\int_0^{D} {\rm d}\epsilon \rho(\epsilon) = \rho_0 D.
\end{align}
Then $\rho(\epsilon)$ is given by
\begin{align}
	\rho(\epsilon)
	 = \frac{d \rho_0}{2} \left( \frac{\epsilon}{D} \right)^{d/2-1},
\label{eq:dos_d}
\end{align}
for each dimension $d$.

For one dimension $d=1$, eq.~(\ref{eq:bound_state_condition}) is evaluated with eq.~(\ref{eq:dos_d}) to yield
\begin{align}
	1 = \frac{\rho_0 U k_{\rm c}}{\kappa} \text{arctan}\frac{k_{\rm c}}{\kappa}
	 \simeq \frac{\pi \rho_0 U k_{\rm c}}{2\kappa},
\end{align}
where we have used the approximation $\text{arctan}(k_{\rm c}/\kappa) \simeq \pi/2$. 
Hence $E_{\rm b}$ is given by
\begin{align}
	E_{\rm b} = -\frac{\pi^2 D}{4} (\rho_0 U)^2.
\label{eq:bound1d}
\end{align}
We note here a resemblance to the bound state appearing in the energy gap of superconductors by magnetic impurities \cite{Soda}. 
Since the DOS of the quasi-particles in superconductors has $\epsilon^{-1/2}$-dependence at the gap edge, the binding energy is given by a formula similar to eq.~(\ref{eq:bound1d}). 

For two dimensions $d=2$, eq.~(\ref{eq:bound_state_condition}) leads to 
\begin{align}
	E_{\rm b} = -D(\text{e}^{1/\rho_0 U}-1)^{-1}.
\label{eq:energy2d}
\end{align}
When $\text{e}^{1/\rho_0 U} \gg 1$ is satisfied, which is a plausible condition in real systems, eq.~(\ref{eq:energy2d}) is reduced to
\begin{align}
	E_{\rm b} = -D \text{e}^{-1/\rho_0 U}.
\label{eq:energy2d_2}
\end{align}
This expression is identical to the well known formulae for the Kondo temperature $T_{\rm K}$ and the superconducting transition temperature. 
It seems that the expression like eq.~(\ref{eq:energy2d_2}) is a common form emerging from the constant DOS.

In the case of three dimensions $d=3$, in contrast to the lower dimensions, there exists a threshold value of $U$. 
A bound state appears only when $U$ is large enough so that $\rho_0 U>1/3$ is satisfied. 
Then $E_{\rm b}$ is given by
\begin{align}
	E_{\rm b} = -\frac{4D}{\pi^2} \left( 1-\frac{1}{3\rho_0 U} \right)^2,
\end{align}
where we have again applied the approximation $\text{arctan}(k_{\rm c}/\kappa) \simeq \pi/2$.

It turns out, from the above discussion, that there may exist one excitonic bound state even though the c-f Coulomb repulsion is on-site. 
As expected, the bound state becomes more stable, the larger the DOS of conduction electrons exists near the gap edge. 
In the semiconducting state, the energy gap weakens the screening effect because of an absence of the particle-hole excitation with an infinitesimal energy transfer. 
Accordingly, the c-f interaction is not necessarily on-site there. 
Hence, there is a possibility that plural bound states exist in semiconducting phase.

\section{Self-Consistent Perturbation Theory}
\label{sec:scpt}
In constructing an approximate scheme by a partial resummation of perturbation series, it is important to let the scheme preserve the natural characteristics such as the analyticity and the sum-rule. 
It has been shown for the system of electron gas that approximate schemes constructed from generating functionals $\Phi$ of the skeleton diagrams conserve the proper characteristics\cite{Baym-Kadanoff}.
Such an approach is also available for the resolvnent method, which deals with the restricted Hilbert space \cite{nca1}.
In this section, we derive equations for the resolvent and some dynamical quantities within the conserving approximation, following the analysis of the vertex part in \S\ref{sec:vf}. 
The c-f Coulomb interaction $U_{\rm cf}$ is considered to be independent of the wave number, since the $\delta$-function potential has turned out to give a bound state in the previous section.

\subsection{Self-energy part}
The self-energy part is obtained by functional differentiation of $\Phi$ with respect to $R_{\gamma}(z)$. 
Diagram in Fig.~\ref{fig:phi} shows the generating functional which leads to the self-energy parts including the ladder-type intermediate-state interactions, Figs.~\ref{fig:vertex1} and \ref{fig:vertex0}. 
The first term generates the NCA self-energy, and others modify it so as to incorporate terms leading to the excitonic bound state. 
\begin{figure}[t]
	\begin{center}
	\includegraphics[width=8.0cm]{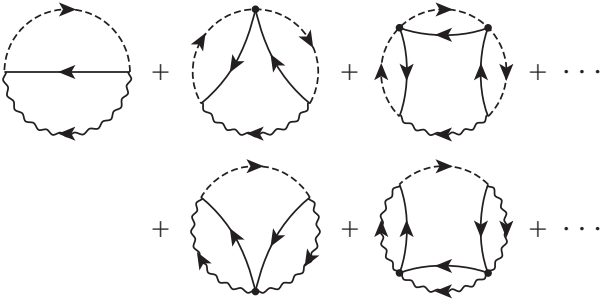}
	\end{center}
	\caption{Diagrams representing the generating functional $\Phi$ which includes the consecutive scattering due to the c-f Coulomb interaction. The first term represents the NCA. }
	\label{fig:phi}
\end{figure}

Generating functional in Fig.~\ref{fig:phi} are reminiscent of those introduced in extensions for the finite-$U$ Anderson model, where the vertex correction incorporates vertual excitations to 4f$^2$ states \cite{Sakai, Kroha, Otsuki_f2v}. 
We can observe a structural correspondence between both generating functionals. 
Accordingly, the knowledge of the schemes concerning the finite-$U$ Anderson model are applicable in this model. 

The self-energy parts generated from $\Phi$ in Fig.~\ref{fig:phi} are diagrammatically represented in Fig.~\ref{fig:self}. 
\begin{figure}[t]
	\begin{center}
	\includegraphics[width=8.5cm]{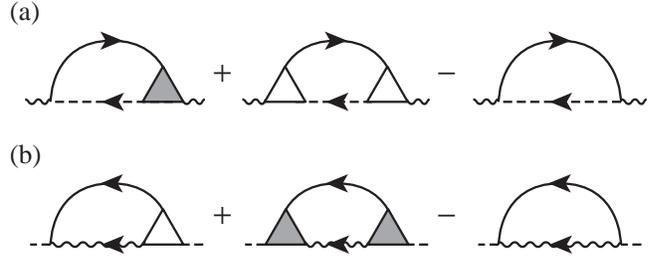}
	\end{center}
	\caption{Diagrammatical representation of the self-energy part (a) $\Sigma_{\alpha}(z)$ and (b) $\Sigma_{\beta}(z)$ in the conserving approximation. }
	\label{fig:self}
\end{figure}
The first term is the ladder-type diagram introduced in Figs.~\ref{fig:vertex1} and \ref{fig:vertex0}. 
In addition, the second diagram is necessary for the conserving approximation. 
The third diagram subtracts a double counting part. 
We introduce auxiliary functions $P_{\alpha}(z)$ and $P_{\beta}(z)$ defined by
\begin{subequations}
\begin{align}
	P_{\alpha}(z) &= \int {\rm d}\epsilon \rho(\epsilon) f(-\epsilon) R_{\alpha} (z-\epsilon), 
\label{eq:P_n}\\
	P_{\beta}(z) &= \int {\rm d}\epsilon \rho(\epsilon) f(\epsilon) R_{\beta} (z+\epsilon).
\label{eq:P_n+1}
\end{align}
\end{subequations}
Since now the c-f Coulomb interaction is on-site, the vertex function $\Lambda_{n+1} (\mib{k}, z)$ and $\Lambda_{n}(\mib{k}, z)$ introduced  respectively in eqs.~(\ref{eq:vertex1}) and (\ref{eq:vertex0}) become independent of the wave vector $\mib{k}$. 
The integral equations are consequently solvable and lead to 
\begin{subequations}
\begin{align}
	\Lambda_{\alpha}(z) &= [1+U_0 P_{\alpha}(z)]^{-1}, \label{eq:vertex_n}\\
	\Lambda_{\beta}(z) &= [1+U_1 P_{\beta}(z)]^{-1}. \label{eq:vertex_n+1}
\end{align}
\end{subequations}
It is clear that the vertex part $\Lambda_{\gamma}(z)$ reduces to unity if the c-f Coulomb interaction does not exist.
The self-energy part $\Sigma_{\alpha}(z)$ of 4f$^n$ state $|\alpha \rangle$ is given with use of $P_{\beta}(z)$ and $\Lambda_{\beta}(z)$ by
\begin{align}
	\Sigma_{\alpha}(z) &= \sum_{\beta} W_{\alpha \beta} \bigg\{
	 P_{\beta}(z) [\Lambda_{\beta}(z)-1] \nonumber \\
	 &\quad + \int {\rm d}\epsilon \rho(\epsilon) f(\epsilon) \tilde{R}_{\beta \alpha}(z+\epsilon) \bigg\}, 
\label{eq:self_alpha} \\
	\tilde{R}_{\beta \alpha}(z) &= R_{\beta}(z) \Lambda_{\alpha}^2(z),
\label{eq:R_beta_alpha}
\end{align}
where $W_{\alpha \beta}$ is defined by $W_{\alpha \beta} = V^2 \sum_m |\langle \beta |f^{\dag}_m |\alpha \rangle|^2$. 
In an analogous way, the self-energy part $\Sigma_{\beta}$ of the 4f$^{n+1}$ state $|\beta \rangle$ is given by
\begin{align}
	\Sigma_{\beta}(z) &= \sum_{\alpha} W_{\alpha \beta} \bigg\{
	 P_{\alpha}(z) [\Lambda_{\alpha}(z)-1] \nonumber \\
	 &\quad + \int {\rm d}\epsilon \rho(\epsilon) f(-\epsilon) \tilde{R}_{\alpha \beta} (z-\epsilon) \bigg\}, \label{eq:self_beta} \\
	\tilde{R}_{\alpha \beta}(z) &= R_{\alpha}(z) \Lambda_{\beta}^2(z).
\label{eq:R_alpha_beta}
\end{align}
These equations are solved numerically combined with eq.~(\ref{eq:resolv}).

We define spectral function $\eta_{\gamma}(\epsilon)$ of the resolvent and that of the defect propagator $\xi_{\gamma}(\epsilon)$. 
They are expressed in terms of the resolvent as\cite{nca3}
\begin{align}
	\eta_{\gamma}(\epsilon) &= -\pi^{-1} \text{Im} R_{\gamma}(\epsilon^+), \\
	\xi_{\gamma}(\epsilon) &= Z_{\rm f}^{-1} {\rm e}^{-\beta \epsilon} (-\pi^{-1}) \text{Im} R_{\gamma}(\epsilon^+), 
	\label{eq:xi}
\end{align}
where $\epsilon^+$ denotes $\epsilon$ plus an infinitesimal complex number.
$\eta_{\gamma}(\epsilon)$ stands for intensity of an addition of the state $|\gamma\rangle$, while $\xi_{\gamma}(\epsilon)$ a removal of the state $|\gamma \rangle$. 
Although $\xi_{\gamma}(\epsilon)$ analytically relates with the resolvent in eq.~(\ref{eq:xi}), the Boltzmann factor makes it difficult to compute $\xi_{\gamma}(\epsilon)$ at low temperatures. 
To avoid the difficulty, we transform the integral equations for the resolvent into a form convenient for the numerical calculation of $\xi_{\gamma}(\epsilon)$ (see Appendix A).
With use of these spectral functions, dynamical quantities are represented without the Boltzmann factor. 
$\xi_{\gamma}(\epsilon)$ is further used to evaluate the partition function $Z_{\rm f}$ in eq.~(\ref{eq:part_func}). 
Thermodynamic quantities such as the entropy and the specific heat are evaluated from $T$-derivatives of the spectral functions, which can be computed with use of another set of integral equations\cite{Otsuki_thermo}.

\subsection{Single-particle Green function}
We now derive the single-particle Green function from the generating functional $\Phi$ in Fig.~\ref{fig:phi}.
The 4f Green function is defined by
\begin{align}
	G_{{\rm f},m} ({\rm i}\epsilon_n)
	&= -\int_0^\beta {\rm d}\tau \langle f_m^{\rm H}(\tau) f_m^{\dag} \rangle
	 {\rm e}^{{\rm i} \epsilon_n \tau} \nonumber \\
	&= \sum_{\alpha \beta} |\langle \beta | f_m^{\dag} |\alpha \rangle |^2
	 G_{\alpha \beta} ({\rm i} \epsilon_n), 
\end{align}
where $\epsilon_n =(2n+1)\pi T$ is the fermion Matsubara frequency and the superscript H indicates the Heisenberg picture. 
$G_{\alpha \beta} ({\rm i}\epsilon_n)$ is defined in terms of the $X$-operators by
\begin{align}
	G_{\alpha \beta} ({\rm i}\epsilon_n)
	= -\int_0^\beta {\rm d}\tau \langle X_{\alpha \beta}^{\rm H} (\tau) X_{\beta \alpha} \rangle
	 {\rm e}^{{\rm i} \epsilon_n \tau}.
\end{align}

We consider a response to an infinitesimal external field which changes the 4f number. 
As far as hybridization process is concerned, our approximation is equivalent to the NCA, since the vertex parts merely describe interactions between the intermediate states. 
Therefore, the formula of the 4f Green function in the NCA is applicable in this case. 
Hence $G_{\alpha \beta}({\rm i}\epsilon_n)$ is given by
\begin{align}
	G_{\alpha \beta} ({\rm i}\epsilon_n)
	 = \frac{1}{Z_{\rm f}} \int_{\rm C} \frac{{\rm d}z}{2\pi {\rm i}} \text{e}^{-\beta z}
	 R_{\alpha} (z) R_{\beta} (z+{\rm i}\epsilon_n),
\end{align}
which is expressed by diagram in Fig.~\ref{fig:green_func}. 
\begin{figure}[t]
	\begin{center}
	\includegraphics[width=3.5cm]{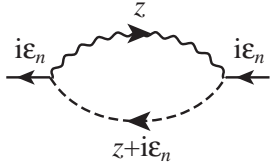}
	\end{center}
	\caption{Diagrammatical representation of the 4f electron Green function. }
	\label{fig:green_func}
\end{figure}
Performing analytic continuation ${\rm i}\epsilon_n = \omega^+$ to real frequencies, we obtain the DOS $\rho_{\alpha \beta}(\omega) = -\pi^{-1} \text{Im} G_{\alpha \beta}(\omega^+)$ of 4f electrons
\begin{align}
	\rho_{\alpha \beta}(\omega) = \int {\rm d}\epsilon [ \xi_{\alpha}(\epsilon) \eta_{\beta}(\epsilon+\omega)
	 + \eta_{\alpha}(\epsilon) \xi_{\beta}(\epsilon+\omega)].
\end{align}

\subsection{Two-particle correlation functions}
We proceed to the two-particle correlation function. 
In general, 4f electrons possess the orbital degrees of freedom. 
Correspondingly there are various kinds of correlation functions other than spin-spin and charge-charge ones. 
Therefore it is reasonable to consider formula for correlation functions in a general form with four subscripts. 
At the end of this subsection, we shall apply it to the dynamical magnetic susceptibility in a model for Ce compounds.

\subsubsection{General formula}
Let us consider a correlation function of operators $A$ and $B$ which do not change the 4f number, i.e.,
 $\langle \alpha|A| \beta \rangle =\langle \alpha|B| \beta \rangle=0$. 
The 4f response function is defined by
\begin{align}
	\chi_{AB} ({\rm i}\nu_n) &= \int_0^{\beta} {\rm d}\tau \langle T_{\tau} A^{\rm H}(\tau) B \rangle
	 {\rm e}^{{\rm i}\nu_n \tau} \nonumber \\
	&= \sum_{\gamma \gamma' \gamma'' \gamma'''}
	 \langle \gamma |A| \gamma' \rangle \langle \gamma'' |B| \gamma''' \rangle
	 \chi_{\gamma \gamma', \gamma'' \gamma '''} ({\rm i}\nu_n),
	\label{eq:def_correlation}
\end{align}
where $\nu_n = 2n\pi T$ denotes the boson Matsubara frequency and $\chi_{\gamma \gamma', \gamma'' \gamma '''} ({\rm i}\nu_n)$ is defined in terms of the $X$-operators by
\begin{align}
	\chi_{\gamma \gamma', \gamma'' \gamma '''} ({\rm i}\nu_n) 
	= \int_0^{\beta} {\rm d}\tau \langle T_{\tau} X_{\gamma \gamma'}^{\rm H} (\tau) X_{\gamma'' \gamma'''} \rangle
	{\rm e}^{{\rm i}\nu_n \tau}.
\end{align}

The NCA requires the vertex correction in the two-particle correlation function if, for example, both 4f$^n$ and 4f$^{n+1}$ states have the magnetic moment\cite{nca4} or the hybridization is anisotropic\cite{Otsuki_st}. 
In the present model, an approximation consistent with $\Phi$ in Fig.~\ref{fig:phi} needs the vertex correction. 
The 4f response function is given by
\begin{align}
	\chi_{\gamma \gamma', \gamma'' \gamma'''}({\rm i}\nu_n)
	= - \frac{1}{Z_{\rm f}} \int_{\rm C} \frac{{\rm d}z}{2\pi {\rm i}} \text{e}^{-\beta z}
	 \Pi_{\gamma \gamma', \gamma'' \gamma'''} (z, z+{\rm i}\nu_n).
\label{eq:suscep}
\end{align}
$\Pi_{\gamma \gamma', \gamma'' \gamma'''}(z, z')$ satisfies the following Bethe-Salpeter-type integral equation:
\begin{align}
	&\Pi_{\gamma \gamma', \gamma'' \gamma'''}(z, z')
	= R_{\gamma}(z) R_{\gamma'}(z') \bigg\{ \delta_{\gamma \gamma'''} \delta_{\gamma' \gamma''} \nonumber \\
	 &+ \sum_{\lambda \lambda'} \int {\rm d}\epsilon
	 \Gamma_{\gamma \gamma', \lambda \lambda'}(z,z'; \epsilon)
	 \Pi_{\lambda' \lambda, \gamma'' \gamma'''}(z+\epsilon, z'+\epsilon) \bigg\}, 
\label{eq:Pi}
\end{align}
where sum of $\lambda$ and $\lambda'$ are taken for all 4f states. 
$\Gamma_{\gamma \gamma', \gamma'' \gamma'''}(z, z'; \epsilon)$ denotes an irreducible vertex function.
Diagrammatical representations of eqs.~(\ref{eq:suscep}) and (\ref{eq:Pi}) are shown in Fig.~\ref{fig:suscep}. 
\begin{figure}[t]
	\begin{center}
	\includegraphics[width=8.5cm]{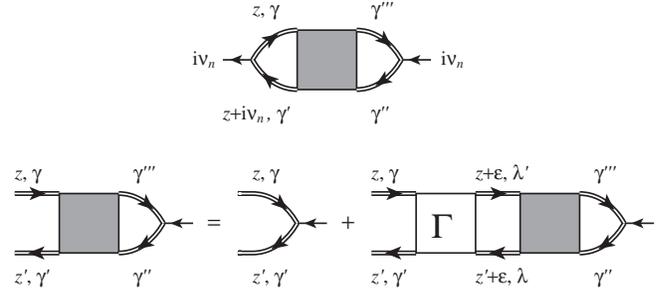}
	\\[0.5cm]
	\end{center}
	\caption{Diagram of the tow-particle correlation function. Double line means all 4f states $\gamma$ including 4f$^n$ and 4f$^{n+1}$, namely, it can be thought of as dashed or wavy line. }
	\label{fig:suscep}
\end{figure}

We classify $\Pi_{\gamma \gamma', \gamma'' \gamma'''}$ by their subscripts, 
i.e., we make use of subscript $\alpha$ (4f$^n$ states) and $\beta$ (4f$^{n+1}$ states) instead of $\gamma$ (whole states). 
It is sufficient to introduce four classes of functions: 
$\Pi^{(10)}_{\beta \beta', \alpha \alpha'}$, 
$\Pi^{(01)}_{\alpha \alpha', \beta \beta'}$, 
$\Pi^{(11)}_{\beta \beta', \beta'' \beta'''}$, and 
$\Pi^{(00)}_{\alpha \alpha', \alpha'' \alpha'''}$.
Analogously, the irreducible vertex functions $\Gamma_{\gamma \gamma', \gamma'' \gamma'''}$ are classified into four parts:
$\Gamma^{(10)}_{\beta \beta', \alpha \alpha'}$, 
$\Gamma^{(01)}_{\alpha \alpha', \beta \beta'}$, 
$\Gamma^{(11)}_{\beta \beta', \beta'' \beta'''}$, and 
$\Gamma^{(00)}_{\alpha \alpha', \alpha'' \alpha'''}$.
In the present approximation, each function is given by
\begin{subequations}
\begin{align}
	&\Gamma^{(10)}_{\beta \beta', \alpha \alpha'} (z, z'; -\epsilon)
	= W_{\alpha \alpha', \beta \beta'} \rho(\epsilon) f(-\epsilon) \nonumber \\
	&\times [\Lambda_{\beta} (z-\epsilon) \Lambda_{\beta'} (z'-\epsilon)
	 +\Lambda_{\alpha'}(z) \Lambda_{\alpha} (z') -1], \\
	&\Gamma^{(01)}_{\alpha \alpha', \beta \beta'} (z, z'; \epsilon)
	= W_{\alpha \alpha', \beta \beta'} \rho(\epsilon) f(\epsilon) \nonumber \\
	&\times [\Lambda_{\beta'} (z) \Lambda_{\beta} (z')
	 +\Lambda_{\alpha}(z+\epsilon) \Lambda_{\alpha'} (z'+\epsilon) -1], \\
	&\Gamma^{(11)}_{\beta \beta', \beta' \beta}  (z, z'; \epsilon) \nonumber \\
	&= -U_1 \int {\rm d}\epsilon' \rho(\epsilon') f(-\epsilon')
	 \rho(\epsilon + \epsilon') f(\epsilon + \epsilon') \nonumber \\
\label{eq:gamma11}
	&\quad \times \sum_{\alpha} [W_{\alpha \beta}
	\tilde{R}_{\alpha \beta}(z-\epsilon') \Lambda_{\beta'}(z'-\epsilon') \nonumber \\
	 &\quad \qquad + W_{\alpha \beta'} \tilde{R}_{\alpha \beta'}(z'-\epsilon') \Lambda_{\beta}(z-\epsilon') ], \\
	&\Gamma^{(00)}_{\alpha \alpha', \alpha' \alpha}  (z, z'; -\epsilon)\nonumber \\
	&= -U_0 \int {\rm d}\epsilon' \rho(\epsilon') f(\epsilon')
	 \rho(\epsilon + \epsilon') f(-\epsilon - \epsilon') \nonumber \\
\label{eq:gamma00}
	&\quad \times \sum_{\beta} [W_{\alpha \beta}
	\tilde{R}_{\beta \alpha}(z+\epsilon') \Lambda_{\alpha'}(z'+\epsilon') \nonumber \\
	 &\quad \qquad + W_{\alpha' \beta} \tilde{R}_{\beta \alpha'}(z'+\epsilon') \Lambda_{\alpha}(z+\epsilon') ],
\end{align}
\end{subequations}
where we have introduced $W_{\alpha \alpha', \beta \beta'}=V^2 \sum_m \langle \beta' |f_m^{\dag}| \alpha \rangle \langle \alpha' | f_m |\beta \rangle$.
The above quantities are diagrammatically expressed in Fig.~\ref{fig:gamma}(a). 
\begin{figure}[t]
	\begin{center}
	\includegraphics[width=8.5cm]{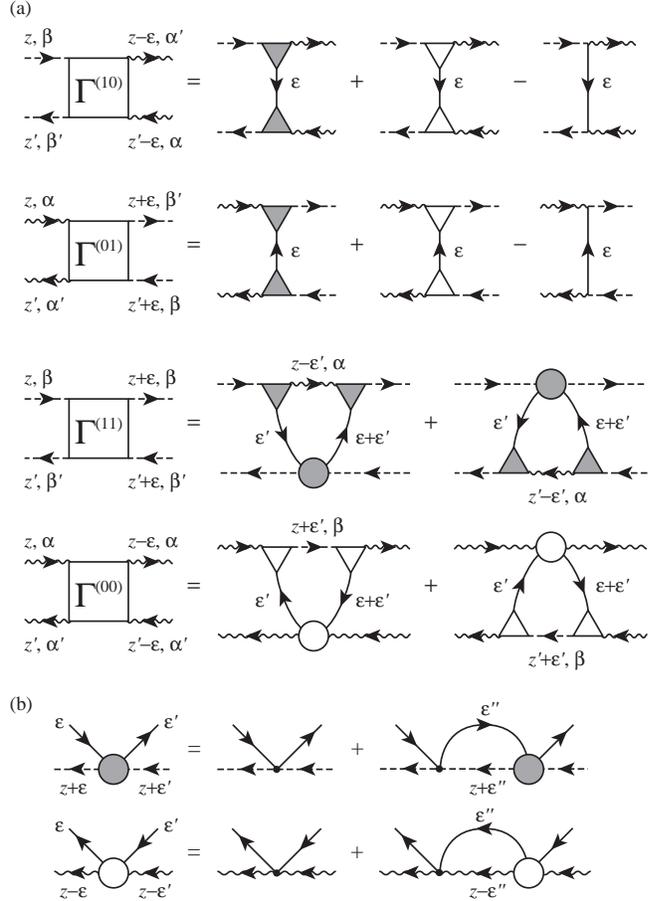}
	\end{center}
	\caption{Diagrammatical representations of (a) the irreducible vertex parts $\Gamma^{(10)}_{\beta \beta', \alpha \alpha'}$, $\Gamma^{(01)}_{\alpha \alpha', \beta \beta'}$, $\Gamma^{(11)}_{\beta \beta', \beta' \beta}$ and $\Gamma^{(00)}_{\alpha \alpha', \alpha' \alpha}$, and (b) equations for the effective c-f Coulomb interactions $\tilde{U}_{\beta}(z)$ and $\tilde{U}_{\alpha}(z)$. }
	\label{fig:gamma}
\end{figure}
The first and second irreducible vertices are what should be considered in the NCA, namely, they do not become zero even if $U_{\rm cf}=0$. 
On the other hand, the third and fourth ones are entirely due to the c-f Coulomb interaction, that is, they dissappear if $U_{\rm cf}=0$. 
In deriving eqs.~(\ref{eq:gamma11}) and (\ref{eq:gamma00}), 
we have introduced effective c-f Coulomb interactions $\tilde{U}_{\beta}(z)$ and $\tilde{U}_{\alpha}(z)$ corresponding to diagrams in Fig.~\ref{fig:gamma}(b). 
Since the c-f Coulomb interaction is now on-site, the integral equations are solvable and give $\tilde{U}_{\beta}(z)=U_1 \Lambda_{\beta}(z)$ and $\tilde{U}_{\alpha}(z)=-U_0 \Lambda_{\alpha}(z)$.

After analytic continuation ${\rm i}\nu_n = \omega^+$ to the real axis in eq.~(\ref{eq:suscep}), 
we obtain an expression for the magnetic spectrum
\begin{align}
	&\text{Im} \chi_{\gamma \gamma', \gamma'' \gamma'''}(\omega^+)
	= (1-{\rm e}^{-\beta \omega}) \frac{1}{Z_{\rm f}} \int \frac{{\rm d}\epsilon}{2 \pi}
	{\rm e}^{-\beta \epsilon} \nonumber \\
	&\times \text{Re} [
	\Pi_{\gamma \gamma', \gamma'' \gamma'''}(\epsilon^-, \epsilon+\omega^+)
	- \Pi_{\gamma \gamma', \gamma'' \gamma'''}(\epsilon^+, \epsilon+\omega^+) ].
\label{eq:Im_chi}
\end{align}
The static susceptibility $\chi(0)$ is given by
\begin{align}
	\chi_{\gamma \gamma', \gamma'' \gamma'''}(0)
	= \frac{1}{Z_{\rm f}} \int \frac{{\rm d}\epsilon}{\pi} {\rm e}^{-\beta \epsilon}
	\text{Im} \Pi_{\gamma \gamma', \gamma'' \gamma'''}(\epsilon^+, \epsilon^+).
\label{eq:chi0}
\end{align}

\subsubsection{Dynamical magnetic susceptibility in a model for Ce compounds}

For the systems of 4f$^0$-4f$^1$, only $\Pi^{(11)}$ contributes to the dynamical magnetic susceptibility since 4f$^0$ has no magnetic moment. 
In the absence of magnetic field, contributions from terms including $\Gamma^{(10)}$ and $\Gamma^{(01)}$ as irreducible vertices become zero, since sum of the magnetic moment appears in such terms.
Hence, $\Gamma^{(11)}$ is the only irreducible vertex relevant to the magnetic susceptibility. 
Dynamical magnetic susceptibility $\chi_{\rm M}({\rm i}\nu_n)$ is then given by
\begin{align}
	\chi_{\rm M}({\rm i}\nu_n)
	= -\frac{N_1 C}{Z_{\rm f}} \int_{\rm C} \frac{{\rm d}z}{2\pi {\rm i}} \text{e}^{-\beta z} \Pi_{\rm M} (z, z+{\rm i}\nu_n),
\end{align}
where $C$ is the Curie constant and $N_1$ is the number of degeneracy of 4f$^1$ states. 
$\Pi_{\rm M}(z, z')$ satisfies the following integral equation:
\begin{align}
	&\Pi_{\rm M}(z, z') = R_1(z) R_1(z') \bigg\{ 1
	- V^2 U_1 \int {\rm d}\epsilon \rho(\epsilon) f(-\epsilon) \nonumber \\
	&\times [ \tilde{R}_{01}(z-\epsilon) \Lambda_1 (z'-\epsilon)
	  +\tilde{R}_{01}(z'-\epsilon) \Lambda_1 (z-\epsilon) ]\nonumber \\
	&\times \int {\rm d}\epsilon' \rho(\epsilon') f(\epsilon') 
	\Pi_{\rm M}(z-\epsilon+\epsilon', z'-\epsilon+\epsilon') \bigg\},
\label{eq:Ce_Pi}
\end{align}
which follows from eqs.~(\ref{eq:Pi}) and (\ref{eq:gamma11}). 
Imaginary part of $\chi_{\rm M}(\omega)$ and the static magnetic susceptibility $\chi_{\rm M}(0)$ are evaluated by combined with eqs.~(\ref{eq:Im_chi}) and (\ref{eq:chi0}), respectively. 
In computing these equations including the Boltzmann factor, we transform eq.~(\ref{eq:Ce_Pi}) into suitable forms (see Appendix B).

\section{Numerical Results}
\label{sec:results}
In this section, we show numerical results of the self-consistent equations developed in the preceding section.
We consider a model for Ce compounds, where 4f state fluctuates between 4f$^0$ and 4f$^1$. 
Degeneracy of 4f$^1$ states is assumed to be six ($j=5/2$), and we do not consider the crystalline electric field splitting.
In this model, all matrix elements of $f_m$ are unity. 
We take a constant DOS of conduction electrons with energy gap
\begin{align}
	\rho(\epsilon) = \frac{1}{2D} \theta(D+E_{\rm g}-|\epsilon|)\theta(|\epsilon| - E_{\rm g}).
\end{align}
We take $D$ as a unit of energy, i.e., $D=1$.

Numerical calculations are first performed for parameters where 4f electron is almost localized, to which the Ce ions usually correspond. 
The case where 4f states are almost unoccupied will be examined as well for comparison.

\subsection{4f$^1$-dominant case}

As a preliminary to systems with energy gap, we show results for the metallic phase to examine the influence of $U_{\rm cf}$ on the Kondo effect.
\begin{figure}[t]
	\begin{center}
	\includegraphics[width=8.5cm]{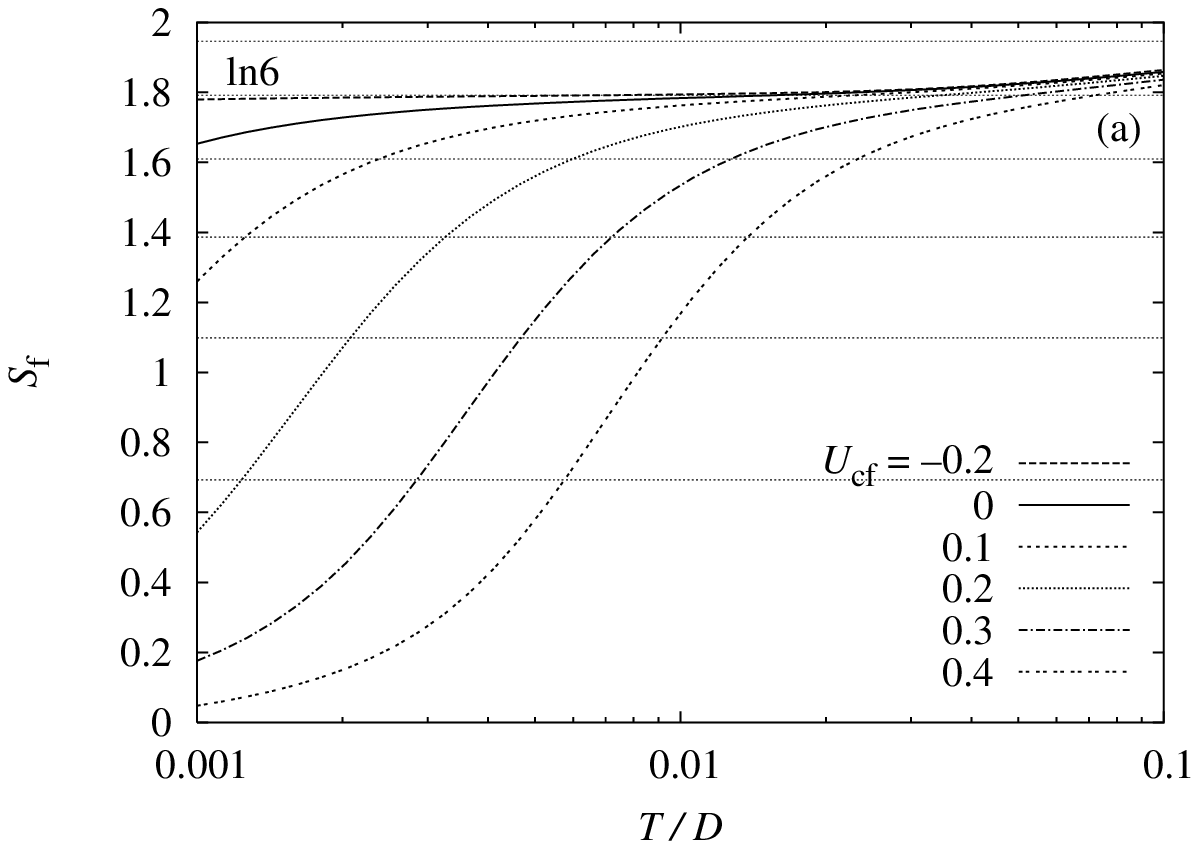}
	\includegraphics[width=8.5cm]{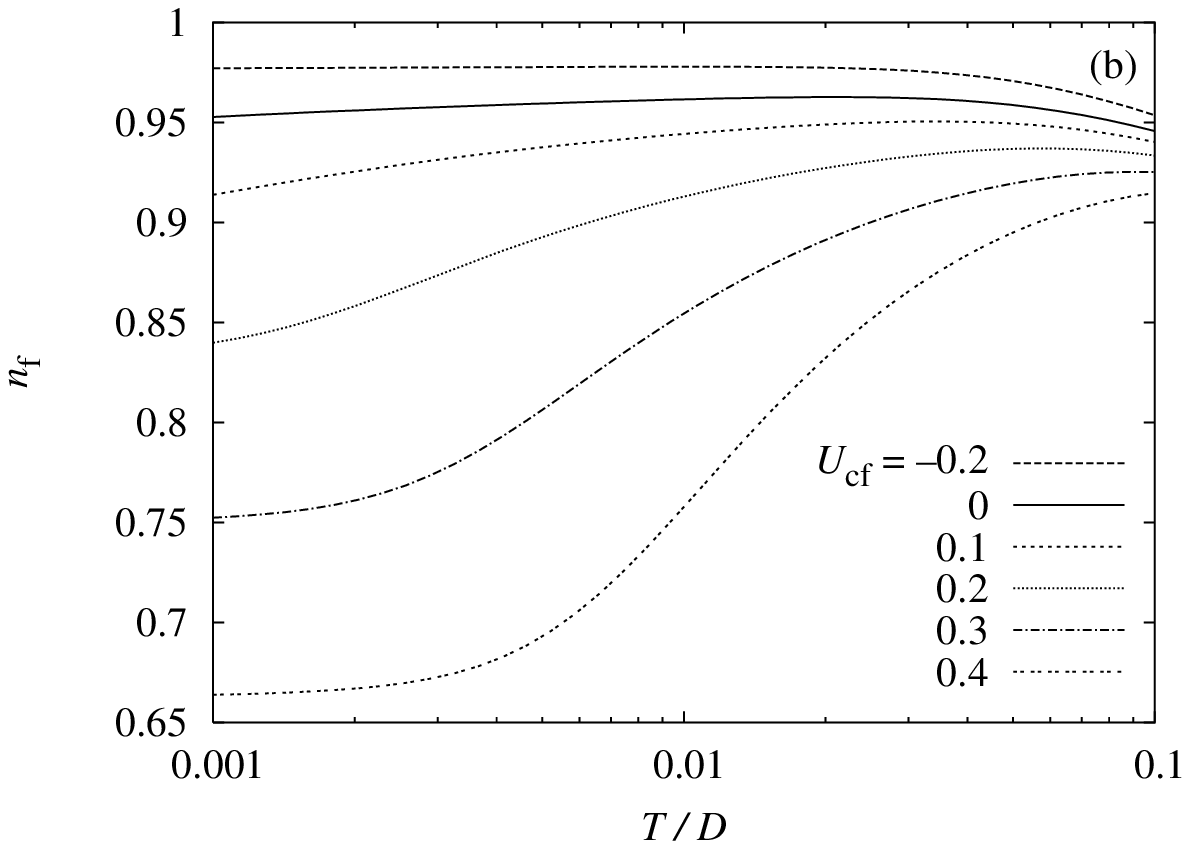}
	\end{center}
	\caption{
	Temperature dependences of (a) the entropy $S_{\rm f}$ and (b) the 4f number $n_{\rm f}$ for several values of $U_{\rm cf}$ ($\nu=1$) without the energy gap in the conduction band. Other parameters are $\epsilon_{\rm f}=-0.2$ and $V=0.1$. }
	\label{fig:gap0_entropy}
\end{figure}
Figure~\ref{fig:gap0_entropy} shows temperature dependences of the entropy $S_{\rm f}$ and the 4f number $n_{\rm f}$ for several values of $U_{\rm cf}$ with $\epsilon_{\rm f}=E_1 -E_0=-0.2$ and $V=0.1$. 
The entropy has been computed without numerical differentiation by solving integral equations for $T$-derivatives of the resolvent\cite{Otsuki_thermo}.
It turns out, from Fig.~\ref{fig:gap0_entropy}(a), that the repulsive values of $U_{\rm cf}$ increase the Kondo temperature. 
The fact means that the effect of $U_{\rm cf}$ is, at least in the metallic phase, simply regarded as a increase of $V$. 
Correspondingly, the 4f number decreases with increasing $U_{\rm cf}$ as shown in Fig.~\ref{fig:gap0_entropy}(b). 
\begin{figure}[t]
	\begin{center}
	\includegraphics[width=8.5cm]{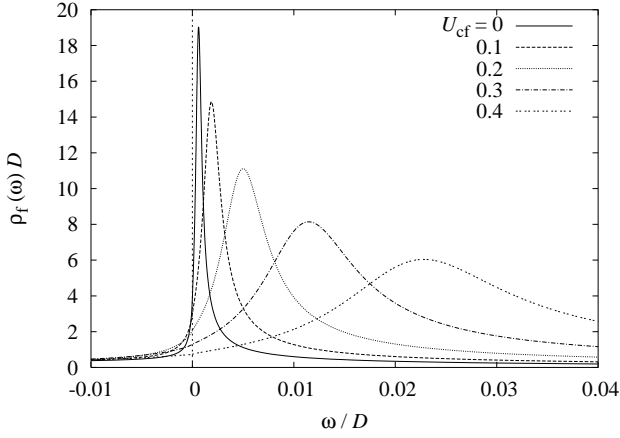}
	\end{center}
	\caption{The DOS $\rho_{\rm f}(\omega)$ of 4f electrons near the Fermi level.  Temperatures are much lower than $T_{\rm K}$ and other parameters are same as in Fig.~\ref{fig:gap0_entropy}. }
	\label{fig:gap0_dos}
\end{figure}
The DOS $\rho_{\rm f}(\omega)$ of 4f electrons near the Fermi level is shown in Fig.~\ref{fig:gap0_dos}. 
The Kondo resonance shifts toward the higher energy indicating increase of the Kondo temperature. 
These results are consistent with the previous study by means of the NRG\cite{Costi-Hewson}.

\begin{figure}[t]
	\begin{center}
	\includegraphics[width=8.5cm]{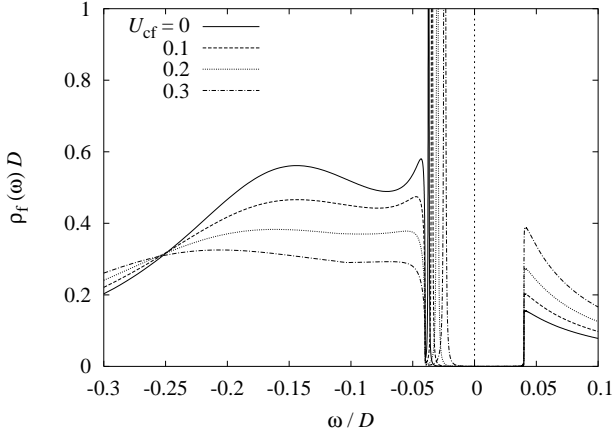}
	\includegraphics[width=8.5cm]{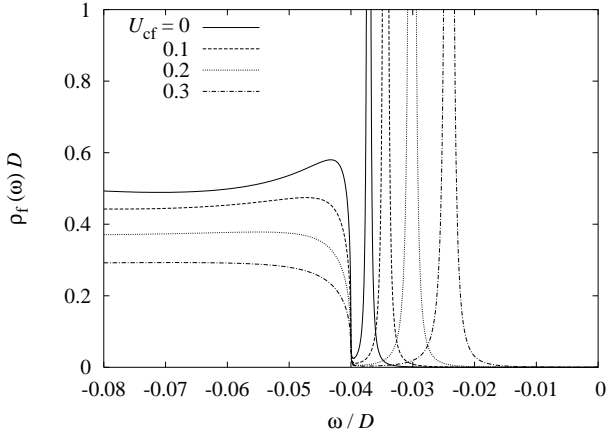}
	\end{center}
	\caption{The DOS $\rho_{\rm f}(\omega)$ for several values of $U_{\rm cf}$ with $\nu=1$. Other parameters are $\epsilon_{\rm f}=-0.2$, $V=0.1$, $E_{\rm g}=0.04$ and $T=0.01$. }
	\label{fig:dos_f}
\end{figure}
We now proceed to the case of semiconducting phase. 
Figure~\ref{fig:dos_f} shows the DOS $\rho_{\rm f}(\omega)=-\pi^{-1} \text{Im}G_{\rm f}(\omega^+)$ of 4f electrons for $E_{\rm g}=0.04$.  
We vary $U_{\rm cf}$ with other parameters fixed at $\epsilon_{\rm f}=-0.2$, $V=0.1$, $\nu=1$ and $T=0.01$. 
In these parameter sets, the Kondo temperature is estimated to be much lower than the energy gap. 
We can observe two peaks of completely different features: broad one at about $\epsilon_{\rm f}$ and considerably sharp one of about $T$ in width being inside the energy gap. 
The latter is an excitation from the excitonic bound state discussed in \S\ref{sec:vf}. 
The bound state becomes stable with increasing $U_{\rm cf}$ as expected. 
Correspondingly, a weight of the peak increases. 
We here comment upon the parameter $\nu$. 
In the case where $\nu$ is set to be zero with other parameters fixed, the bound state shifts to upper energy. 
Since now the 4f number is more than 0.9, we recognize that the Hartree term enhances the effect of $U_{\rm cf}$ to yield a larger binding energy.

It is confirmed from Fig.~\ref{fig:dos_f} that the exciton state appears even though $U_{\rm cf}=0$. 
This is because the second order process with respect to hybridization effectively acts as a substitution for the repulsion between 4f and conduction electrons, as mentioned in \S2. 
Furthermore, the large degeneracy $N_1=6$ of 4f states enhances the effective repulsion. 
Namely, the almost localized model yields $N_1 V^2/|\epsilon_{\rm f}|$ as a strength of the effective c-f interaction.
In the corresponding parameters, it is estimated to be $N_1 V^2 /|\epsilon_{\rm f}| =0.3$, and therefore leads to the binding energy $E_{\rm b} \simeq -1.3 \times 10^{-3}$ from eq.~(\ref{eq:energy2d_2}). 
The estimation turns out to be in good agreement with the binding energy measured in Fig.~\ref{fig:dos_f}.

\begin{figure}[t]
	\begin{center}
	\includegraphics[width=8.5cm]{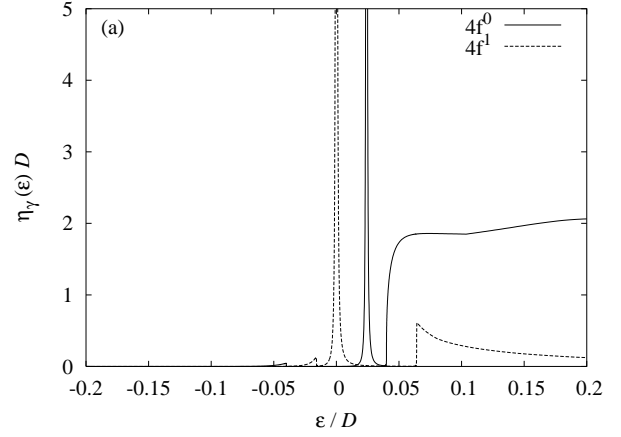}
	\includegraphics[width=8.5cm]{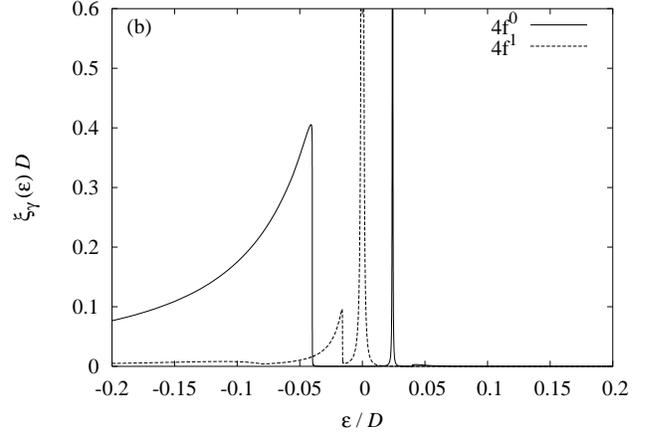}
	\end{center}
	\caption{(a) Spectral functions $\eta_{\gamma}(\epsilon)$ of the resolvents and (b) those of the defect propagators $\xi(\epsilon)$ for $U_{\rm cf}=0.3$ with the same parameters as in Fig.~\ref{fig:dos_f}. An origin of the energy has been set to an arbitrary value (see text).}
	\label{fig:spec_eta_xi}
\end{figure}
In order to see the origin of the bound state, we examine the spectral functions $\eta_{\gamma}(\epsilon)$ of the resolvents and the defect propagators $\xi_{\gamma}(\epsilon)$. 
Figure~\ref{fig:spec_eta_xi} shows the results with $U_{\rm cf}=0.3$. 
In general, $\eta_{\gamma}(\epsilon)$ is finite only above a threshold energy at zero temperature, while $\xi_{\gamma}(\epsilon)$ is finite below the threshold. 
We have adjusted an origin of the energy for the spectral functions so as to set the threshold at $\epsilon \simeq 0$. 
It turns out from Fig.~\ref{fig:spec_eta_xi}(a) that the 4f$^1$ states are located at lower energy than the 4f$^0$ state. 
The fact indicates that the ground state is magnetic, and excludes the possibility of the Kondo effect as an origin of the sharp peak in $\rho_{\rm f}(\omega)$. 
This is also confirmed by the residual entropy of $S_{\rm f}=\ln 6$ and the Curie law behavior of the static magnetic susceptibility (no figure). 
Comparison of $\eta_{\gamma}(\epsilon)$ with $\rho_{\rm f}(\omega)$ in Fig.~\ref{fig:dos_f} indicates that the structure of the single-particle excitation is mostly determined by $\eta_0(\epsilon)$. 
It follows that the bound state comes from the 4f$^0$ resolvent, where the intermediate 4f$^1$ states interact with the holes in valence band by the effective attraction. 
The bound state appearing in $\eta_0 (\epsilon)$ remains in $\xi_0(\epsilon)$ due to the thermal population.

An influence of the impurity on the electrons of the conduction and valence band appears in the site-diagonal part of the Green function.
The local DOS $\rho_{\rm c}(\epsilon)$ at the 4f site is given by
\begin{align}
	\rho_{\rm c}(\omega) = -\frac{1}{\pi} \text{Im} [ \bar{g}(\omega^+) + \bar{g}^2(\omega^+) V^2 G_{\rm f}(\omega^+)],
\label{eq:dos_c}
\end{align}
where $\bar{g}(\omega)$ is the site-diagonal part of the unperturbed Green function for conduction electrons defined by $\bar{g}(\omega^+) = \sum_{\mib{k}} g_{\mib{k}}(\omega^+)/N$.
\begin{figure}[t]
	\begin{center}
	\includegraphics[width=8.5cm]{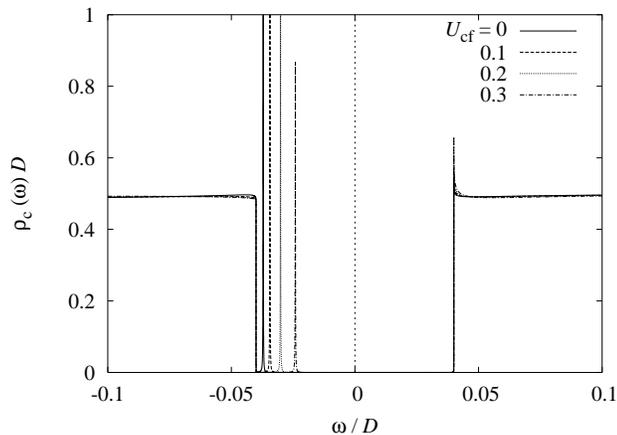}
	\end{center}
	\caption{The DOS $\rho_{\rm c}(\omega)$ of the site-diagonal part of the conduction electrons with the same parameters as in Fig.~\ref{fig:dos_f}.}
	\label{fig:dos_c}
\end{figure}
Figure~\ref{fig:dos_c} shows $\rho_{\rm c}(\omega)$ for the same parameters as in Fig.~\ref{fig:dos_f}. 
We can see an obvious difference from the bare one $\rho(\omega)$. 
Sharp peak is perceived at the same energy as the bound state observed in $\rho_{\rm f}(\omega)$ in Fig.~\ref{fig:dos_f}. 
The state inside the gap is caused by the impurity scattering represented by the second term in eq.~(\ref{eq:dos_c}).

We next examine an influence of the bound state on the magnetic excitations. 
Figure~\ref{fig:mag_spec} shows $\text{Im}\chi_{\rm M}(\omega)$ for the parameters identical with above. 
\begin{figure}[t]
	\begin{center}
	\includegraphics[width=8.5cm]{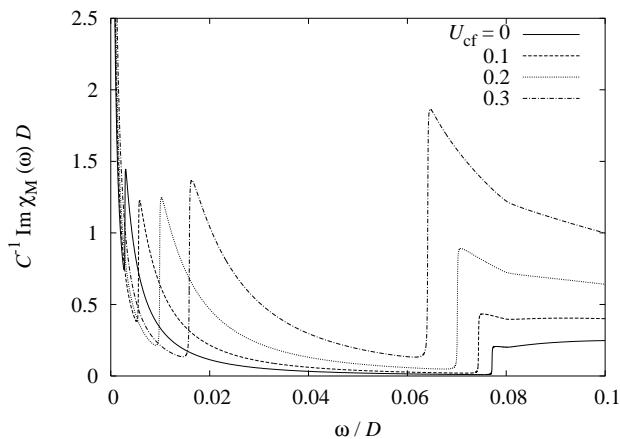}
	\end{center}
	\caption{The magnetic spectrum $\text{Im}\chi_{\rm M} (\omega)$ for the parameters in Fig.~\ref{fig:dos_f}.}
	\label{fig:mag_spec}
\end{figure}
The vertex correction for $\chi_{\rm M}(\omega)$ is not necessary to be calculated since now $\nu=1$ (then $U_1=0$). 
There exist mainly the following three structures. 
One is the quasi-elastic scattering with extremely narrow width. 
This peak is due to the degeneracy of 4f$^1$ states.
In addition, two structures are observed for inelastic excitations. 
The peaks are due to the edges of the gap observed in $\rho_{\rm f}(\omega)$ in Fig.~\ref{fig:dos_f}. 
The excitation energy, or the threshold energy of the excitation, are represented as $E_{\rm g} \pm \omega_{\rm b}$, where $\omega_{\rm b}<0$ denotes an energy where the exciton peak is located in $\rho_{\rm f}(\omega)$. 
$\omega_{\rm b}$ is related to the binding energy $E_{\rm b}$ as $E_{\rm b}=-E_{\rm g}+|\omega_{\rm b}|<0$ for the 4f localized case. 
The upper inelastic peak at $E_{\rm g}+|\omega_{\rm b}|$ corresponds to an excitation from the bound state to unoccupied 4f states. 
On the other hand, the lower one at $E_{\rm g}-|\omega_{\rm b}|$ is due to a transition between the bound state and the occupied states. 

As demonstrated above, the bound state clearly appears in the dynamical responses at low temperatures. 
The next interest is their thermal evolution. 
Figure~\ref{fig:T_depend}(a) shows temperature dependence of $\rho_{\rm f}(\omega)$ for $U_{\rm cf}=0.3$. 
\begin{figure}[t]
	\begin{center}
	\includegraphics[width=8.5cm]{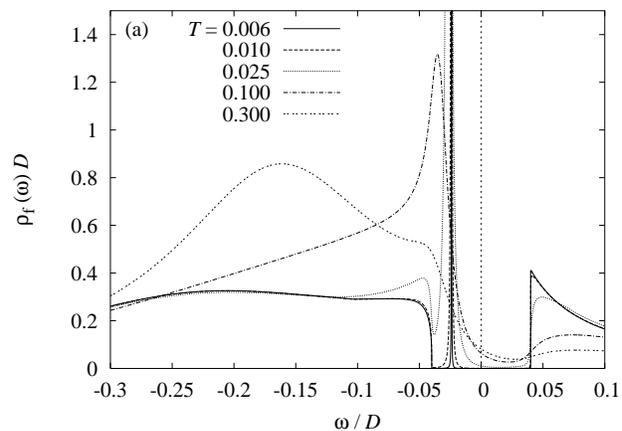}
	\includegraphics[width=8.5cm]{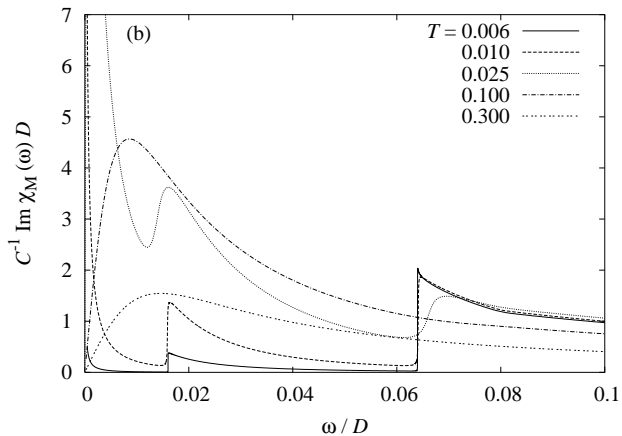}
	\end{center}
	\caption{Temperature dependences of (a) $\rho_{\rm f}(\omega)$ and (b) $\text{Im}\chi_{\rm M}(\omega)$ for $U_{\rm cf}=0.3$. Other parameters are same as in Fig.~\ref{fig:dos_f}.}
	\label{fig:T_depend}
\end{figure}
As temperature increases, the peak of the bound state broadens, and eventurally merges with the broad hill at $T=0.025$, which is close to the binding energy. 
At $T=0.100$ a trace of the bound state is seen at the gap edge of the combined spectrum. 
At higher temperature, the spectrum is almost composed of the excitation from the original 4f$^1$ states. 
\begin{figure}[t]
	\begin{center}
	\includegraphics[width=8.5cm]{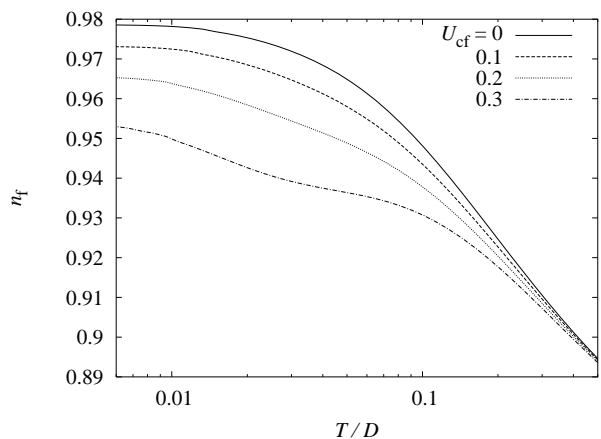}
	\end{center}
	\caption{Temperature dependences of the 4f number $n_{\rm f}$ for the parameters in Fig.~\ref{fig:dos_f}.}
	\label{fig:number}
\end{figure}
Corresponding to the evolution into broadened spectra with increasing temperature, the 4f number $n_{\rm f}$ decreases as shown in Fig.~\ref{fig:number}. 
At low temperatures, $U_{\rm cf}$ reduces $n_{\rm f}$ in a manner similar to the gapless case. 

We next show temperature dependences of $\text{Im}\chi_{\rm M}(\omega)$ in Fig.~\ref{fig:T_depend}(b). 
As for the quasi-elastic peak, it monotonously increases and narrows with decreasing temperature. 
Between the two inelastic peaks at $T=0.010$, the lower one is reduced at lower temperature $T=0.006$, while the upper one keeps the intensity. 
It is conceivable that the lower excitation is enhanced due to the thermal population at temperatures comparable with the energy gap.


\subsection{4f$^0$-dominant case}
We move on to the parameters where 4f states are mostly unoccupied. 
This situation is in fact far from actual Ce ions, but is taken for comparison with the practical model. 
Figure~\ref{fig:f0-dos_f} shows the DOS $\rho_{\rm f}(\omega)$ for $\epsilon_{\rm f}=0.06$ and $T=0.005$. 
\begin{figure}[t]
	\begin{center}
	\includegraphics[width=8.5cm]{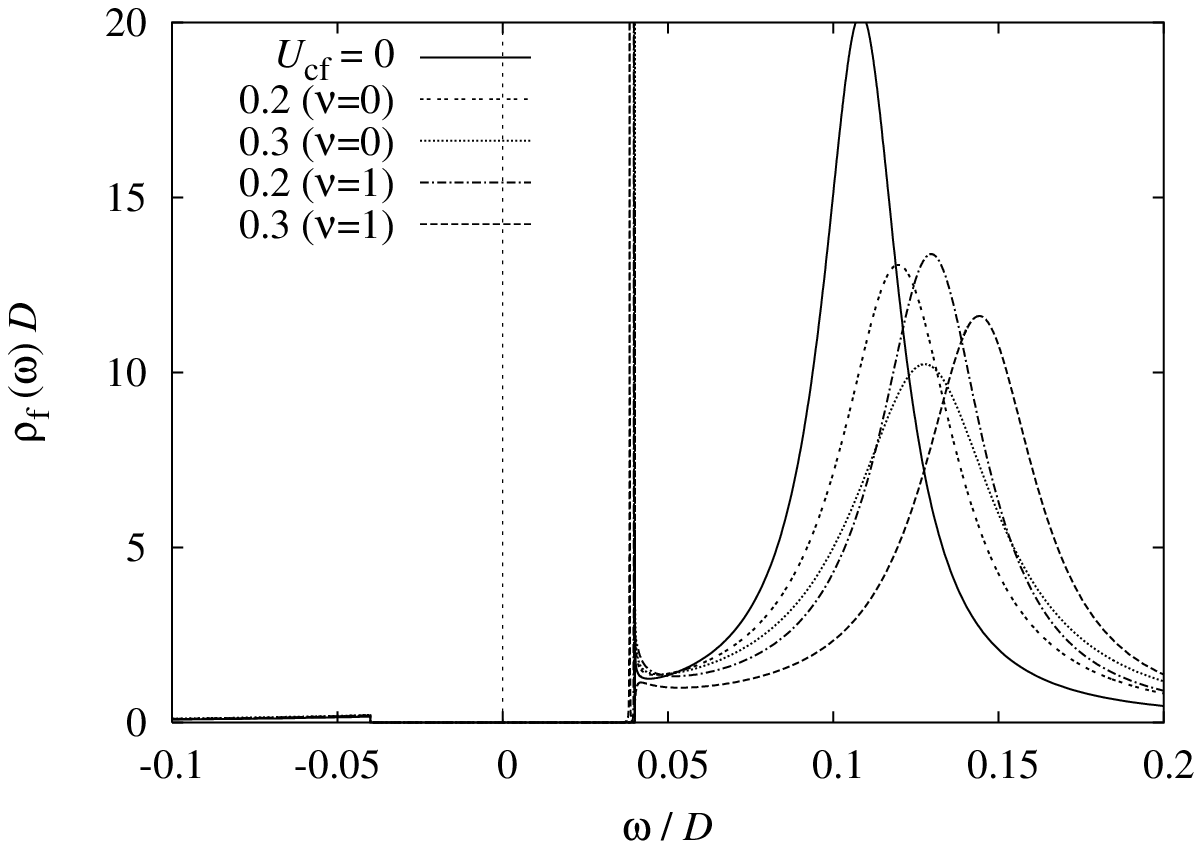}
	\includegraphics[width=8.5cm]{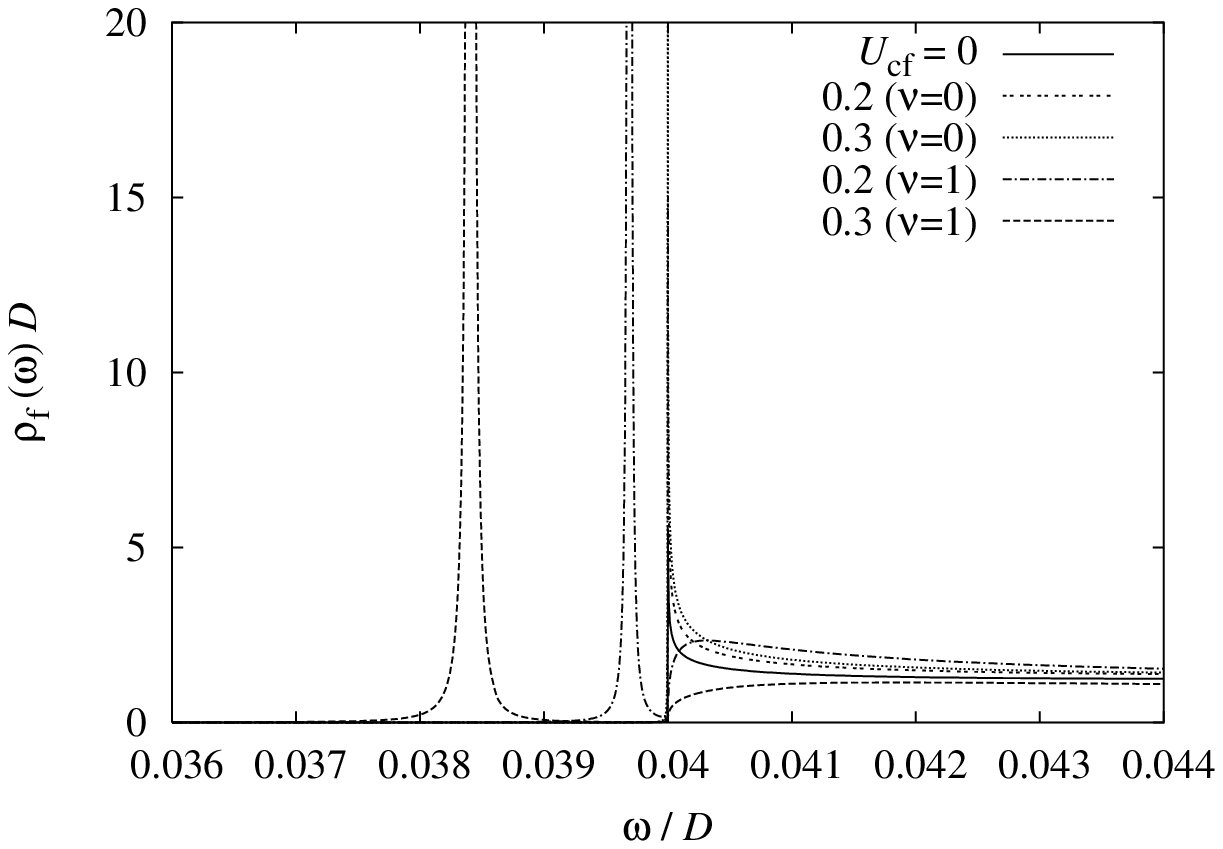}
	\end{center}
	\caption{The DOS $\rho_{\rm f}(\omega)$ for several values of $U_{\rm cf}$ and $\nu$. Other parameters are $\epsilon_{\rm f}=0.06$, $V=0.1$, $E_{\rm g}=0.04$ and $T=0.005$. }
	\label{fig:f0-dos_f}
\end{figure}
Several pairs of values are chosen as parameters $U_{\rm cf}$ and $\nu$. 
In the absence of the term of $U_{\rm cf}$, no distinct bound state can be found in contrast to the 4f localized case. 
Since the degeneracy of 4f states does not work in the second order process of hybridization in this case, the effective c-f interaction remains small. 
For the finite values of $U_{\rm cf}$, we have chosen both $\nu=0$ and $\nu=1$ for comparison.  
In Fig.~\ref{fig:f0-dos_f}, finite values of $U_{\rm cf}$ with $\nu=0$ slightly enhance the binding energy, while larger increases are seen in the case of $\nu=1$.
Since 4f number $n_{\rm f}$ is about 0.15 to 0.20, the Hartree term is largely canceled with $\nu=0$, but not with $\nu=1$. 
It follows that the Hartree term contributes to the effective c-f interaction as in the localized 4f model.

Figure~\ref{fig:f0-mag_spec} shows $\text{Im}\chi_{\rm M}(\omega)$ for the same parameters. 
\begin{figure}[t]
	\begin{center}
	\includegraphics[width=8.5cm]{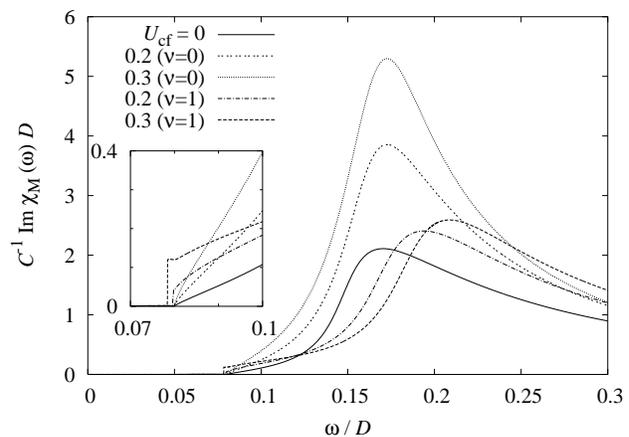}
	\end{center}
	\caption{$\text{Im}\chi_{\rm M}(\omega)$ for the same parameters as in Fig.~\ref{fig:f0-dos_f}. }
	\label{fig:f0-mag_spec}
\end{figure}
The vertex correction has been computed for $\nu=0$, which gives a finite value of $U_1$. 
Unlike the previous case, no quasi-elastic scattering is found since the ground state is a singlet. 
Inelastic scattering occurs above $2E_{\rm g}=0.08$, which correspond to a transition across the energy gap. 
We interpret from the knowledge of $\rho_{\rm f}(\omega)$ in Fig.~\ref{fig:f0-dos_f} that the spectrum reflects a distribution of the unoccupied 4f states shown around $\omega\sim 0.1$ in $\rho_{\rm f}(\omega)$. 
By comparing spectra for $U_{\rm cf}=0$ and $U_{\rm cf}=0.3$ $(\nu=1)$, we conclude that the bound state influences the threshold of the excitation gap. 
The spectrum for $U_{\rm cf}=0.3$ $(\nu=1)$ is discontinuous at the threshold, which shifts to lower energy due to the bound state, while for $U_{\rm cf}=0$ the inelastic spectrum continuously arises at $2E_{\rm g}$. 
% ...

\section{Summary and Discussion}

We have demonstrated an appearance of the excitonic bound state in the semiconducting phase based on the extended Anderson model with the c-f Coulomb interaction.  
The bound state is ascribed to an effective attraction between the 4f electron and valence holes, and between the 4f hole and conduction electrons. 
An importance of these processes is clarified by the perturbation theory which incorporates leading contributions of the c-f Coulomb interaction as a vertex correction to hybridization. 
Our microscopic derivation of the bound state supports previous discussions on the exciton in the mixed-valence semiconductors\cite{Alekseev, Kikoin, Kasuya}.

Compared to the exciton in ordinary semiconductors, 
an important difference is that the bound state can exist at the ground state because the 4f electron mixes with conduction electrons without optical absorption. 
In other words, a source of the excitonic bound states appears as an effective potential in the single-particle Green function, while in the case of the optical absorption a direct attraction in the two-particle response function leads to the bound state. 
Consequently, the excitonic bound state shows up in the 4f DOS as an extra $\delta$-function-like peak located inside the energy gap, in addition to a broad peak corresponding to the original 4f level. 
On the other hand, if the original 4f level is close to the Fermi level, the spectrum becomes single peak structure inside the energy gap. 

In ordinary semiconductors with electron donors or acceptors, the attractive Coulomb potential around the impurities gives rise to the bound state. 
The donor level lies below the conduction band and provides electron carriers by thermal excitations, while the acceptor level lies above the valence band and provides hole carriers. 
On the other hand, in the present model with rare-earth impurities, the bound state located around the 4f site do not contribute to the electrical conduction. 
Namely, the effective c-f attraction simply produces the 4f level inside the energy gap with the chemical potential fixed, on the contrary to the donor or acceptor level.

A source of the bound state in mixed-valence systems is not only the Coulomb interaction but also the effective interaction due to the higher order process of hybridization. 
In fact, it has been demonstrated for the well-localized 4f electron systems that the excitonic bound state appears in the 4f DOS even though $U_{\rm cf}=0$. 
Therefore, the bound state naturally appears around Ce ions in the well-localized 4f systems as long as an energy gap exists. 
In the case where the 4f states are almost unoccupied, however, a distinctive peak by the excitonic bound state does not emerge without $U_{\rm cf}$. 
These contrastive results are well understood by considering each hybridization process as follows. 
When the 4f electron is removed from the localized 4f electron system, the excitonic bound state take places in the resultant 4f$^0$ state, whose self-energy is enhanced due to the $N_1$-fold degeneracy of 4f$^1$ states. 
In the opposite situation, the degeneracy never works in the 4f$^1$-state self-energy.

It is interesting that the energy of the excitonic bound state $|E_{\rm b}|$ and the Kondo temperature $T_{\rm K}$ are given by analogous formulae. 
Suppose 4f electron is well localized and $U_{\rm cf}=0$, both the effective c-f repulsion and the antiferro-exchange coupling are given by $N_1 V^2 / |\epsilon_{\rm f}|$, and therefore $|E_{\rm b}|$ and $T_{\rm K}$ become identical. 
When the system is metallic, the 4f electron forms the Kondo singlet, whose energy gain is given by $T_{\rm K}$ determined from the above coupling constant. 
This is the same situation in semiconductor, provided the energy gap is much smaller than $T_{\rm K}$. 
On the other hand, if there exists a large energy gap compared to $T_{\rm K}$, the excitonic bound state arises at $|E_{\rm b}| \simeq T_{\rm K}$ inside from the gap edge in $\rho_{\rm f}(\omega)$, instead of the Kondo resonance peak. 

It has been shown that the bound state affects magnetic excitations. 
The situations are different depending on whether the gound state has the magnetic moment or not. 
When the system has no magnetic moment at low temperatures, low-energy excitations does not take place. 
If there is no excitonic bound state, excitations gradually start to occur at the gap energy. 
In the case of the presence of the bound state, on the other hand, the threshold of the magnetic excitation becomes discontinuous. 
An energy of the threshold is lower than the original gap due to the existence of the bound state. 
For the magnetic ground state, the quasi-elastic scattering appears, and as in the singlet ground state the excitonic bound state make the spectrum discontinuous at the threshold. 


We have devoted ourselves to the single-impurity model in this paper. 
The system corresponds to such a situation where magnetic impurities like Ce are doped into semiconductors.
On the other hand, present results can contribute to study on the exciton in the periodic systems as well. 
Suppose the excitonic bound state is formed at each site, it transfers from site to site through conduction electrons. 
Propagation to other site may lead to, for instance, broadening of the local bound states and wave-number dependence of the intensity and the binding energy. 
The situation becomes more complicated in the Kondo semiconductors, where 4f electrons themselves take part in a formation of the energy gap. 
It is an interesting issue whether 4f electrons can simultaneously play roles both in an evolution of the hybridization gap and in the appearance of the excitonic bound state. 
The periodic systems can be dealt with by the present theory combined with the dynamical mean field theory. 
They are left for future studies. 

\section*{Acknowledgment}
The author would like to acknowledge valuable discussions with Professor Y. Kuramoto and Dr. H. Kusunose. 
The author is supported by Research Fellowships of the Japan Society for the Promotion of Science for Young Scientists.

\appendix
\section{Equations for the Defect Propagator}
The spectral function of the defect propagator $\xi_{\gamma}(\epsilon)$ relates with the resolvent analytically in eq.~(\ref{eq:xi}). 
However, the Boltzmann factor makes it difficult to perform direct computations at low temperatures. 
In order to obtain them by numerical calculations, we transform the whole integral equations for the resolvent, following ref.~\citen{nca3}. 
We introduce an operator $\mathcal{P}^{(\xi)}$ defined by 
$\mathcal{P}^{(\xi)}R(\epsilon) = -Z_{\rm f}^{-1} \text{e}^{-\beta z} \pi^{-1} \text{Im} R(\epsilon^+)$. 
The operator $\mathcal{P}^{(\xi)}$ converts eq.~(\ref{eq:resolv}) into
\begin{align}
	\xi_{\gamma}(\epsilon) = |R_{\gamma}(\epsilon^+)|^2 \sigma_{\gamma}(\epsilon), 
\end{align}
where $\sigma_{\gamma}(\epsilon)$ denotes $\sigma_{\gamma}(\epsilon) = \mathcal{P}^{(\xi)}\Sigma_{\gamma}(\epsilon)$. 
Operating $\mathcal{P}^{(\xi)}$ on eqs.~(\ref{eq:self_alpha}) and (\ref{eq:self_beta}), we obtain
\begin{align}
	\sigma_{\alpha}(\epsilon)
	&= \sum_{\beta} W_{\alpha \beta} \bigg\{ P_{\beta}^*(\epsilon) \lambda_{\beta}(\epsilon)
	 + \pi_{\beta}(\epsilon) [\Lambda_{\beta}(\epsilon)-1] \nonumber \\
	 &\quad + \int {\rm d}\epsilon' \rho(\epsilon') f(-\epsilon') \tilde{\xi}_{\beta \alpha}(\epsilon+\epsilon') \bigg\}, \\
	\sigma_{\beta}(\epsilon)
	&= \sum_{\alpha} W_{\alpha \beta} \bigg\{ P_{\alpha}^*(\epsilon) \lambda_{\alpha}(\epsilon)
	 + \pi_{\alpha}(\epsilon) [\Lambda_{\alpha}(\epsilon)-1] \nonumber \\
	 &\quad + \int {\rm d}\epsilon' \rho(\epsilon') f(\epsilon') \tilde{\xi}_{\alpha \beta}(\epsilon-\epsilon') \bigg\}. 
\end{align}
Here $\pi_{\gamma}(\epsilon)=\mathcal{P}^{(\xi)} P_{\gamma}(\epsilon)$ and $\lambda_{\gamma}(\epsilon)=\mathcal{P}^{(\xi)} \Lambda_{\gamma}(\epsilon)$ are given by
\begin{align}
	\pi_{\alpha}(\epsilon)
	&= \int {\rm d}\epsilon' \rho(\epsilon') f(\epsilon') \xi_{\alpha}(\epsilon-\epsilon'), \\
	\pi_{\beta}(\epsilon)
	&= \int {\rm d}\epsilon' \rho(\epsilon') f(-\epsilon') \xi_{\beta}(\epsilon+\epsilon'), \\ 
	\lambda_{\alpha}(\epsilon) &= -U_0 |\Lambda_{\alpha}(\epsilon^+)|^2 \pi_{\alpha}(\epsilon), \\
	\lambda_{\beta}(\epsilon) &= -U_1 |\Lambda_{\beta}(\epsilon^+)|^2 \pi_{\beta}(\epsilon),
\end{align}
which follow from eqs.~(\ref{eq:P_n}), (\ref{eq:P_n+1}), (\ref{eq:vertex_n}) and (\ref{eq:vertex_n+1}), respectively. 
Equations~(\ref{eq:R_beta_alpha}) and (\ref{eq:R_alpha_beta}) give expressions for $\tilde{\xi}_{\beta \alpha}(\epsilon)=\mathcal{P}^{(\xi)}\tilde{R}_{\beta \alpha}(\epsilon)$ and $\tilde{\xi}_{\alpha \beta}(\epsilon)=\mathcal{P}^{(\xi)}\tilde{R}_{\alpha \beta}(\epsilon)$ as
\begin{align}
	\tilde{\xi}_{\beta \alpha}(\epsilon) &= \xi_{\beta}(\epsilon) \text{Re} [\Lambda_{\alpha}^2(\epsilon^+)] \nonumber \\
	 &\quad + 2 \text{Re}[R_{\beta}(\epsilon^+)] \text{Re}[\Lambda_{\alpha}(\epsilon^+)] \lambda_{\alpha}(\epsilon), \\
	\tilde{\xi}_{\alpha \beta}(\epsilon) &= \xi_{\alpha}(\epsilon) \text{Re} [\Lambda_{\beta}^2(\epsilon^+)] \nonumber \\
	 &\quad + 2 \text{Re}[R_{\alpha}(\epsilon^+)] \text{Re}[\Lambda_{\beta}(\epsilon^+)] \lambda_{\beta}(\epsilon).
\end{align}
Norms of the spectral functions of the defect propagator, namely the partition function $Z_{\rm f}$, are determined by the following sum-rule:
\begin{align}
	\int {\rm d}\epsilon \sum_{\gamma} \xi_{\gamma}(\epsilon) =1. 
\end{align}
This equation follows from eq.~(\ref{eq:part_func}).

\section{Equations for the Magnetic Susceptibility}
The magnetic excitation and the static magnetic susceptibility is given by eqs.~(\ref{eq:Im_chi}) and (\ref{eq:chi0}) in the forms including the Boltzmann factor, respectively. 
For numerical calculations, we rewrite eq.~(\ref{eq:Ce_Pi}) as well as eqs.~(\ref{eq:Im_chi}) and (\ref{eq:chi0}), following ref.~\citen{nca4}. 
We introduce, for each value of $\omega$, simplified notations as follows:
\begin{align}
	p(\epsilon) &= \Pi_{\rm M}(\epsilon^+, \epsilon+\omega^+),
	 \nonumber \\
	\tilde{p}(\epsilon) &= \Pi_{\rm M}(\epsilon^-, \epsilon+\omega^+),
	 \nonumber \\
	r(\epsilon) &= R_1(\epsilon^+) R_1(\epsilon+\omega^+),
	 \nonumber \\
	\tilde{r}(\epsilon) &= R_1(\epsilon^-) R_1(\epsilon+\omega^+),
	 \nonumber \\
	q(\epsilon) &= \tilde{R}_{01}(\epsilon^+) \Lambda_1(\epsilon+\omega^+)
	 + \tilde{R}_{01}(\epsilon +\omega^+) \Lambda_1(\epsilon^+),
	 \nonumber \\
	\tilde{q}(\epsilon) &= \tilde{R}_{01}(\epsilon^-) \Lambda_1(\epsilon+\omega^+)
	 + \tilde{R}_{01}(\epsilon +\omega^+) \Lambda_1(\epsilon^-).
\end{align}
The argument $\omega$ has been omitted in the left-hand side of each definitions. 
Equation~(\ref{eq:Ce_Pi}) is rewritten with use of the above symbols as
\begin{align}
	p(\epsilon) = r(\epsilon) \bigg\{
	 1 -& V^2 U_1 \int {\rm d}\epsilon'' \rho(\epsilon'') f(-\epsilon'')
	q(\epsilon-\epsilon'') \nonumber \\
	&\times \int {\rm d}\epsilon' \rho(\epsilon') f(\epsilon') p(\epsilon+\epsilon'-\epsilon'') \bigg\}.
	\label{eq:p}
\end{align}
We further introduce an operator $\mathcal{P}$ defined by 
\begin{align}
	\mathcal{P}p(\epsilon) = \frac{1}{2\pi Z_{\rm f}} {\rm e}^{-\beta \epsilon}
	 [\tilde{p}(\epsilon) - p(\epsilon)].
\end{align}
With use of the operator $\mathcal{P}$, the imaginary part of the magnetic susceptibility is given by
\begin{align}
	\text{Im} \chi_{\rm M}(\omega^+)
	= N_1 C (1-{\rm e}^{-\beta \omega}) \int {\rm d}\epsilon \text{Re} \mathcal{P}p(\epsilon).
\label{eq:Im_chi_M}
\end{align}
We derive an integral equation for the $\mathcal{P}p(\epsilon)$.
Operating $\mathcal{P}$ on eq.~(\ref{eq:p}), we obtain
\begin{align}
	\mathcal{P} p(\epsilon) &= p(\epsilon) [\mathcal{P}r(\epsilon)] / r(\epsilon) \nonumber \\
	&\quad - V^2 U_1 \tilde{r}(\epsilon) \int {\rm d}\epsilon \rho(\epsilon'') f(\epsilon'')
	 [\mathcal{P} q(\epsilon-\epsilon'')] \nonumber \\
	&\qquad \times \int {\rm d}\epsilon' \rho(\epsilon') f(\epsilon') p(\epsilon+\epsilon'-\epsilon'')
	 \nonumber \\
	&\quad - V^2 U_1 \tilde{r}(\epsilon) \int {\rm d}\epsilon \rho(\epsilon'') f(\epsilon'')
	 \tilde{q}(\epsilon-\epsilon'') \nonumber \\
	&\qquad \times \int {\rm d}\epsilon' \rho(\epsilon') f(-\epsilon') \mathcal{P} p(\epsilon+\epsilon'-\epsilon'').
\label{eq:P_p}
\end{align}
As an initial condition of this equation, $\mathcal{P}r(\epsilon)$ and $\mathcal{P}q(\epsilon)$ are evaluated from the original quantities as follows:
\begin{align}
	\mathcal{P}r(\epsilon) &= \text{i}
	 \xi_1(\epsilon) R_1 (\epsilon+\omega^+), \nonumber \\
	\mathcal{P}q(\epsilon) &= \text{i}
	 \tilde{\xi}_{01}(\epsilon) \Lambda_1 (\epsilon+\omega^+)
	 + \text{i} \tilde{R}_{01} (\epsilon+\omega^+) \lambda_1(\epsilon).
\end{align}
In the first step, $p(\epsilon)$ is computed by the numerical iteration of eq.~(\ref{eq:p}) for a certain value of $\omega$. 
Then the results are substituted into eq.~(\ref{eq:P_p}) to compute $\mathcal{P}p(\epsilon)$, which leads to the dynamical magnetic susceptibility through eq.~(\ref{eq:Im_chi_M}). 
This procedure is iterated for each value of $\omega$.

For the static magnetic susceptibility, we introduce, instead of $\mathcal{P}$, another operator $\hat{\mathcal{P}}$ defined by
\begin{align}
	\hat{\mathcal{P}} p(\epsilon) = \frac{1}{\pi Z_{\rm f}} {\rm e}^{-\beta \epsilon} \text{Im} p(\epsilon).
\end{align}
Then $\chi_{\rm M}(0)$ is evaluated with use of $\hat{\mathcal{P}}$ as
\begin{align}
	\chi_{\rm M}(0) = N_1 C \int {\rm d}\epsilon \hat{\mathcal{P}} p(\epsilon).
\end{align}
The equations for $\hat{\mathcal{P}} p(\epsilon)$ are obtained by replacing $\mathcal{P}$, $\tilde{r}$ and $\tilde{q}$ with $\hat{\mathcal{P}}$, $r^*$ and $q^*$, respectively, in eq.~(\ref{eq:P_p}). 
An initial condition for the equation is given by
\begin{align}
	\hat{\mathcal{P}} r(\epsilon) &= -2 \xi_1(\epsilon) \text{Re} R_1(\epsilon), \nonumber \\
	\hat{\mathcal{P}} q(\epsilon) &=
	 - 2 \tilde{\xi}_{01}(\epsilon) \Lambda_1(\epsilon^+)
	 - 2 \tilde{R}_{01}^* (\epsilon^+) \lambda_1(\epsilon).
\end{align}
$\hat{\mathcal{P}} p(\epsilon)$ is evaluated by combination of eq.~(\ref{eq:p}) with $\omega=0$ and modified one of eq.~(\ref{eq:P_p}).

\end{document}